\documentclass{jfm}
\usepackage{graphicx}
\usepackage{epstopdf, epsfig}
\usepackage{multirow}

\usepackage{changes}
\definechangesauthor[name={Xiao}, color=purple]{hx}
\setremarkmarkup{\textcolor{blue}{(\textit{Comments}: #2 -- #1)}}

\usepackage{graphicx}
\usepackage{amssymb}
\usepackage{lineno}
\usepackage{subfig}
\usepackage[colorinlistoftodos]{todonotes}
\usepackage{mathtools}
\usepackage{booktabs}
\usepackage[numbers]{natbib}
\usepackage{array}
\usepackage{url}
\usepackage[section]{placeins}
\usepackage{bm}
\usepackage[ruled,linesnumbered]{algorithm2e}
\usepackage{gensymb}
\usepackage{lineno}
\usepackage{url}
\usepackage{listings}
\usepackage[colorlinks=true]{hyperref}

\usepackage{changes}
\definechangesauthor[name={Reviewer 1}, color = blue]{R1}
\definechangesauthor[name={Reviewer 2}, color = red]{R2}
\definechangesauthor[name={Reviewer 3}, color = cyan]{R3}
\definechangesauthor[name={All reviewers}, color = purple]{All}
\definechangesauthor[name={Editor}, color = brown]{Editor}
\definechangesauthor[name={Author}, color = olive]{Author}

\newcommand{\mb}[1]{\bm{#1}}

\newcommand{\ms}[1]{\boldsymbol{#1}}
\newcommand{\Cnorm}[1]{\left\Vert #1 \right\Vert_{\Omega}}
\newcommand{\Vnorm}[1]{\left\Vert #1 \right\Vert_{n}}
\newcommand{\Cinner}[1]{\left< #1 \right>_{\Omega}}

\newcommand{\disc}[1]{[#1]}
\newcolumntype{P}[1]{>{\centering\arraybackslash}p{#1}}
\newcolumntype{M}[1]{>{\centering\arraybackslash}m{#1}}
\DeclareMathOperator*{\argmin}{arg\,min}

\shorttitle{Ill-Conditioning of RANS Equations}
\shortauthor{J.-L. Wu, H. Xiao, R. Sun and Q. Wang}

\title{RANS Equations with Explicit Data-Driven Reynolds Stress Closure Can Be Ill-Conditioned}

\author{Jin-Long Wu\aff{1},
 Heng Xiao\aff{1\corresp{\email{hengxiao@vt.edu}}},
 Rui Sun\aff{1}    
 \and
 Qiqi Wang\aff{2}}

\affiliation{\aff{1}Kevin T. Crofton Department of Aerospace and Ocean Engineering, Virginia Tech, Blacksburg, VA 24060, USA
\aff{2}Department of Aeronautics and Astronautics, MIT, Cambridge, MA 02139, USA}

\begin{document}

\maketitle

\begin{abstract}
Reynolds-averaged Navier--Stokes (RANS) simulations with turbulence closure models continue to play important roles in industrial flow simulations. However, the commonly used linear eddy viscosity models are intrinsically unable to handle flows with non-equilibrium turbulence (e.g., flows with massive separation). Reynolds stress models, on the other hand, are plagued by their lack of robustness. Recent studies in plane channel flows found that even substituting Reynolds stresses with errors below 0.5\%  from direct numerical simulation (DNS) databases into RANS equations leads to velocities with large errors (up to 35\%). While such an observation may have only marginal relevance to traditional Reynolds stress models, it is disturbing for the recently emerging data-driven models that treat the Reynolds stress as an explicit source term in the RANS equations, as it suggests that the RANS equations with such models can be ill-conditioned.  So far, a rigorous analysis of the condition of such models is still lacking. As such, in this work we propose a metric based on local condition number function for \textit{a priori} evaluation of the conditioning of the RANS equations. We further show that the ill-conditioning cannot be explained by the global matrix condition number of the discretized RANS equations.  Comprehensive numerical tests are performed on turbulent channel flows at various Reynolds numbers and additionally on two complex flows, i.e., flow over periodic hills and flow in a square duct. Results suggest that the proposed metric can adequately explain observations in previous studies, i.e., deteriorated model conditioning with increasing Reynolds number and better conditioning of the implicit treatment of Reynolds stress compared to the explicit treatment. This metric can play critical roles in the future development of data-driven turbulence models by enforcing the conditioning as a requirement on these models.
\end{abstract}


\section{Introduction}
Reynolds-averaged Navier--Stokes (RANS) simulations play an important role in industrial simulations of turbulent flows.
The state-of-the-art eddy-viscosity models (e.g., $k$--$\varepsilon$, $k$--$\omega$ and S--A models~\citep{launder74application,wilcox88reassessment,menter94two-equation,spalart92one}) are based on the assumption that the turbulence production and dissipation are in equilibrium, and thus they perform poorly in flows with non-equilibrium turbulence~\citep{speziale96towards,hamlington08reynolds,hamlington14modeling}, e.g., flows with massive separations or abrupt mean flow changes. On the other hand, Reynolds stress models (RSM), also referred to as second moment closures, take into account the transport of Reynolds stresses and thus have better performance than eddy viscosity models for flows with non-equilibrium turbulence~\citep{pope00turbulent}. The CFD Vision 2030 white paper of NASA identified advanced turbulence modeling based on Reynolds stresses models as a priority for aeronautic technological advancement in the coming decades~\citep{slotnick14cfd}.  However, so far Reynolds stress models have not been widely used in the engineering applications despite their theoretical superiority. A key shortcoming of the Reynolds stress models is their lack of stability and robustness, i.e., the difficulty in achieving convergence~\citep{basara03new,maduta17improved} and the high sensitivity to the modeling of unclosed terms (particularly the pressure--rate-of-strain tensor) in the Reynolds stress transport equations.

Among the leading causes of numerical instability of Reynolds stress models is the intricate coupling as dictated by the Reynolds stress transport equations~\citep[][Chap.~7]{pope00turbulent}. First, the shear stress $\tau_{xy}$ is responsible for the production of streamwise fluctuations $\tau_{xx}$, where $x$ and $y$ denote streamwise and wall-normal coordinates, respectively. Then, the turbulent kinetic energy in $\tau_{xx}$ is redistributed by the pressure--rate-of-strain to the other two normal components of the Reynolds stress tensor including $\tau_{yy}$, which in turn generates the turbulent shear stress $\tau_{xy}$ through interactions with the mean strain field.
Another possible cause for the numerical instability is the ill-conditioning of the RANS momentum equations themselves. Here \emph{model conditioning} refers to the sensitivity of the solved quantities (e.g., mean velocity and pressure fields) to the modeled terms (Reynolds stresses). In computational fluid dynamics codes that solve RANS equations and turbulence transport equations in a segregated manner, ill-conditioned systems lead to numerical instability due to the sensitivity of solved velocities to residuals in the Reynolds stress from iteration to iteration. While the chain-coupling mechanism described above is well-known in the turbulence modeling literature, the conditioning of RANS equations has rarely been mentioned. This is probably due to the fact that the two causes are closely intertwined in traditional models based on Reynolds stress transport equations and the former dominated.

\subsection{Unique challenges in data-driven Reynolds stress closures}
Recently, data-driven turbulence modeling has emerged as a promising field of research with a number of approaches been proposed in the past few years~\cite[e.g.,][]{ling16reynolds,singh16using,wang17physics-informed,weatheritt2016novel}. Of particular relevance to the present discussions are the data-driven Reynolds stress closures~\citep{ling16reynolds,wang17physics-informed,geneva2018quantifying} in which the Reynolds stresses are obtained directly from machine learning models trained on high-fidelity simulation databases without solving partial differential equations (PDEs) or using explicit algebraic models.
In such models, the chain coupling mechanism related to the Reynolds stress transport equations as described above is not a relevant cause of numerical instability, and thus the possible ill-conditioning of RANS equations is placed under spotlight. 
Moreover, these data-driven Reynolds stress models do not have explicit expressions for the Reynolds stress~\citep{ling16reynolds,wang17physics-informed}, which make it difficult to treat the Reynolds stresses implicitly in the RANS equations to improve model conditioning.
For example, \cite{wang17physics-informed} used random forests regression to predict Reynolds stresses based on direct numerical simulation (DNS) databases and reported that the solved mean velocity field does not improve over the original RANS simulations, even though the predicted Reynolds stresses showed significant improvements. Such a paradox clearly demonstrates the gap between \textit{a priori} and \textit{a posteriori} performances in turbulence models based on data-driven Reynolds stress closures. That is, an ill-conditioned model can amplify small \textit{a priori} errors in the modeled terms to large \textit{a posteriori} errors in the solved quantities, which defeats all efforts in the improvement of the closure models. 
Similar gaps have been observed in data-driven subgrid-scale (SGS) stress models for large eddy simulations (LES). For example, \cite{gamahara17searching} reported that their neural network model predicted better SGS stresses than the Smagorinsky model for turbulent channel flows, but the mean velocity predictions were much less satisfactory.
Given the large number of efforts in developing data-driven Reynolds stresses model ~\citep[e.g.,][]{ling16reynolds,wang17physics-informed,zhang2018machine} and subgrid-scale models~\citep[e.g.,][]{vollant2017subgrid,maulik2017neural,maulik2018sub,king2016autonomic} for turbulent flows, the conditioning in such models must be closely examined
along with their \textit{a priori} predictive performances.
Note, however, that some data-driven approaches, e.g., those based on correcting turbulence transport equations~\citep{parish16paradigm,singh2017machine} or discovering analytical turbulent constitutive relations by using symbolic regression~\citep{weatheritt2016novel} are less affected by the conditioning discussed here.

\subsection{DNS data as the ideal scenario for data-driven Reynolds stress closures}
Using Reynolds stress obtained from DNS data to solve the RANS equations for velocity can be considered the ideal scenario for data-driven Reynolds stress closure, at least for those without analytical expressions.
Solving for mean velocities with a given Reynolds stress field is referred to as ``propagation'' in this work, i.e., propagation of Reynolds stresses to mean velocities by solving the RANS equations.
Such a methodology represents an absolute upper limit of performances for data-driven Reynolds stress models as pursued by~\cite{ling16reynolds}, \cite{wang17physics-informed}, and \cite{zhang2018machine}. It allows us to concentrate on the conditioning of this class of machine-learning-based Reynolds stress models without considering the \textit{a priori} performance of any specific model.

Somewhat surprisingly, even solving RANS equations with Reynolds stresses from DNS data can lead to large errors in the velocities, which has been demonstrated in the recent work of~\cite{thompson16methodology}. They propagated Reynolds stresses obtained from DNS to mean velocities for turbulent channel flows at a wide range of Reynolds numbers. These DNS were performed with extreme caution by reputable groups~\citep{del2003spectra,bernardini2014velocity,lee15direct}, and it was verified that the errors in the reported Reynolds stresses were indeed very small, typically less than $0.5\%$ as shown in Table~\ref{tab:summary}. \cite{thompson16methodology} showed that the solved mean velocity has an unsatisfactory agreement with the DNS data at high frictional Reynolds numbers (e.g., $Re_{\tau}=5200$). \cite{poroseva16on} also confirmed such observations.
To motivate our work, we reproduced the two studies of solving for mean velocities by using the DNS Reynolds stresses obtained from \cite{lee15direct}, and the results are summarized in Table~\ref{tab:summary}.  Although the solved mean velocities shown here are accurate up to $Re_{\tau}=1000$, researchers have reported that large errors in the propagated velocities can be found at Reynolds number as low as $Re_{\tau}=395$ depending on the DNS data used~\citep{poroseva18personal}. 

The percentage errors shown in Table~\ref{tab:summary} are computed by comparing
the DNS Reynolds stresses and the  mean velocities propagated therefrom against their respective truths, which are obtained according to \cite{thompson16methodology}.
Specifically, the true velocities are taken as the DNS mean velocities, and the true Reynolds shear stresses are computed based on the following analytical solution:
\begin{linenomath}
\begin{equation*}
\tau=\left(1-\frac{y}{h}\right)\tau_w-\rho\nu\frac{dU}{dy}
\end{equation*}
\end{linenomath}
from the DNS mean velocities and wall shear stresses $\tau_w$, where $y$ denotes the wall distance ($y=0$ indicates the wall) and $h$ denotes half channel width; $\rho$ and $\nu$ represent density and kinematic viscosity, respectively. 

\begin{table}
  \caption{Summary of the results of the channel flow test, showing percentage errors in the turbulent shear stresses and the propagated mean velocities. The large errors in the high Reynolds number case $\mathrm{Re}_{\tau}=5200$ is highlighted. The true Reynolds shear stresses are obtained by analytical integrating the RANS equation with the DNS mean velocities~\citep{thompson16methodology}.}
  \begin{center}
  \begin{tabular}{r|ccccc}
    \hline
    Frictional Reynolds number ($\mathrm{Re}_{\tau}$)  & 180   & 550   & 1000  & 2000  & 5200\\
    \hline \hline
        Error in turbulent shear stresses 
        & & & & & \\
 \emph{volume averaged}   & 0.17\% & 0.21\% & 0.03\% & 0.15\% & 0.31\% \\
    \emph{maximum}
    & 0.43\% & 0.38\% & 0.07\% & 0.23\% & 0.41\% \\
    \hline
      Errors in mean velocities 
      & &  &  &  &  \\
\emph{volume averaged}
    & 0.25\% & 1.61\% & 0.17\% & 2.85\% & \textbf{21.6}\% \\
    \emph{maximum}
    & 0.36\% & 2.70\% & 0.25\% & 5.48\% & \textbf{35.1}\% \\
    \hline
  \end{tabular}
 \end{center}
 \label{tab:summary}
\end{table}

It can be seen in Table~\ref{tab:summary} that the errors in the Reynolds stresses are less than $0.5\%$. The errors in the Reynolds stress can be attributed to different sources, e.g., statistical sampling errors (due to inadequate averaging of the DNS results), iterative errors (due to lack of convergence when solving the linear equations in DNS), and interpolation errors  (when interpolating the DNS data to the RANS mesh).  All these errors are inevitable in any DNS data.
It is noted that errors in the Reynolds stresses vary non-monotonically with respect to the Reynolds number.  Such a coincidental trend should not be overly or literally interpreted, since the data shown here are a result of complex interactions that lead to superposition and cancellation among various sources as discussed above.  It suffices to point out that all the DNS were obtained with extreme caution, as is evident from the very small errors in the Reynolds stresses.  Nevertheless, the errors in the mean velocity solved from the Reynolds stresses clearly increase monotonically with the Reynolds number, because they are dominated by the conditioning of RANS equations as we will show later.  Specifically, while the errors in the solved mean velocity are also less than $0.5\%$ at low Reynolds number ($Re_{\tau}=180$),  such errors can be as high as $35\%$ at high Reynolds number ($Re_{\tau}=5200$). This observation suggests that the RANS equations can be ill-conditioned by directly substituting the modeled Reynolds stress into the equations, which is a common practice in current data-driven RANS modeling. Such results raise several critical questions on Reynolds-stress-based data-driven turbulence models. Specifically, 
\begin{itemize}
\item How to explain the deteriorated conditioning (i.e., increased amplification of errors in Reynolds stresses to velocities) with increasing Reynolds numbers observed in these studies? 
\item Is there a quantitative metric that can characterize the conditioning of RANS equations with a given turbulence model? 
\item What are the implications of the above observations to data-driven turbulence models that treat Reynolds stresses as source terms in the RANS equations?
\end{itemize}

\subsection{Summary and contribution of present work}
Our present work aims to answer the above questions by proposing a quantitative metric and elucidating the relevance of the above-mentioned studies to data-driven turbulence modeling. Throughout this study, \emph{no turbulence models are used}. Rather, Reynolds stresses obtained from DNS database are used to represent the ideal performance for any data-driven Reynolds stress closures. A local condition number function based on the work of \cite{chandrasekaran95on} is derived as a metric to assess the conditioning property for turbulence models. Numerical tests on turbulent channel flows at various Reynolds numbers and two more complex flows suggest that the proposed metric can adequately explain observations in previous studies, i.e., deteriorated model conditioning with increasing Reynolds number, and better conditioning of the implicit treatment of Reynolds stress compared to the explicit treatment. As an application of the proposed approach, \cite{wu18data-driven} improved the conditioning of the data-driven Reynolds stress model of \cite{wang17physics-informed} by training machine learning models for the linear and nonlinear parts of Reynolds stress separately. The obtained Reynolds stress model achieved good conditioning and satisfactory predictive accuracy simultaneously.

The rest of this paper is organized as follows. Section~\ref{sec:meth} introduces the global condition number and shows that it fails to explain the deteriorated conditioning with increasing Reynolds numbers. A local condition number function is derived to achieve such an objective. In Section~\ref{sec:results}, the local condition number is used to evaluate the conditioning of RANS equations with data-driven Reynolds stress closures. Section~\ref{sec:discussion} discusses the conditioning of RANS equations with traditional Reynolds stress transport models and the monolithic coupling for both traditional and data-driven models. Finally, conclusions are presented in Section~\ref{sec:conclusion}.

\section{Methodology}
\label{sec:meth}
Consider the steady state Reynolds-averaged Navier--Stokes equations for incompressible, constant density
fluids:
\begin{linenomath}
\begin{eqnarray}
  \label{eq:ns}
  \bm{u} \cdot  \nabla \bm{u} - \nu \nabla^2 \bm{u} 
  + \nabla p - \nabla \cdot \boldsymbol{\tau} & = 0 \\
  \nabla \cdot   \bm{u} & = 0 
\end{eqnarray}
\end{linenomath}
where $\bm{u}$ is the mean flow velocity; $\nu$ is molecular viscosity; $p$ is the pressure normalized by the
constant density of the fluid; $\boldsymbol{\tau}$ is the Reynolds stress tensor, which needs to be modeled. For simplicity we first consider a Reynolds-stress-based model where $\boldsymbol{\tau}$ is obtained by solving a transport equation in a segregated manner with the RANS equations or by a data-driven function~\citep[see e.g.,][]{ling16reynolds}. The objective of this work is to investigate the sensitivity of the obtained mean velocity with respect to small perturbations on the Reynolds stress. 

For notation simplicity, we introduce nonlinear operator $\mathcal{N}$ to include the convection and diffusion terms with
\begin{linenomath}
\begin{equation}
\mathcal{N}(\bm{u}) =
\bm{u} \cdot  \nabla \bm{u} - \nu \nabla^2 \bm{u} 
\end{equation}
\end{linenomath}
The RANS momentum equation in Eq.~(\ref{eq:ns}) can be written as
\begin{linenomath}
\begin{equation}
  \label{eq:ns-concise-N}
  \mathcal{N}(\bm{u}) = \nabla \cdot \boldsymbol{\tau}-\nabla p
\end{equation}
\end{linenomath}
In numerical solvers, the convection term is first linearized around the current velocity $\overline{\mathbf{U}}_{0}$ to obtain 
the linearized RANS equations:
\begin{linenomath}
\begin{equation}
  \label{eq:ns-concise-L}
  \mathcal{L}(\bm{u}) = \nabla \cdot \boldsymbol{\tau}-\nabla p
\end{equation}
\end{linenomath}
where $\mathcal{L}$ is the linearized operator of $\mathcal{N}$, i.e.,
\begin{linenomath}
\begin{equation}
\label{eq:L-definition}
\mathcal{L}(\bm{u}) =
\bm{u}_0 \cdot  \nabla \bm{u} - \nu \nabla^2 \bm{u} 
\end{equation}
\end{linenomath}

The linearized equation~(\ref{eq:ns-concise-L}) is then discretized on a CFD mesh to obtain a linear system of the following form:
\begin{linenomath}
\begin{equation}
  \label{eq:ns-matrix}
  \mathsf{A}  \mathbf{U} = \bm{b}
\end{equation}
\end{linenomath}
where we denoted $\mb{b} = \disc{\nabla\cdot\boldsymbol{\tau} - \nabla p}$ as the imbalance between the two forces, pressure gradient and Reynolds stress divergence; $\mathbf{U} = \disc{\bm{u}}$ is the discretized velocity field to be
solved for. Both $\bm{b}$ and $\mathbf{U}$ are $n \times 1$ vectors, where $n$ is the
number of cells or grid points in the mesh. The matrix $\mathsf{A}$ with dimension $n \times n$
comes from the implicit discretization of the linearized convection term and the molecular
diffusion term. In this work, we focus on the conditioning of linearized RANS equations. This is because most CFD codes deal with the linearized RANS equations in each iteration when solving the nonlinear RANS equations. Therefore, the conditioning metrics studied here are valid within each iteration, even though the flow of concern may deviate from the linearized RANS equations.

\subsection{Matrix-norm as a measure of model conditioning}
\label{sec:global-cn}
We first show the derivation of the traditional matrix-norm-based condition number and explain why it fails to distinguish the sensitivities of solving for mean velocity at different Reynolds numbers as shown in Table~\ref{tab:summary}. Following the definition of matrix norm, the norm of the error in the velocity is bounded as
follows\footnote{As explained in the notation, the norms $\| \overline{\mathbf{U}}\|$, $ \|\bm{b}
  \|$ are taken of the discretized vectors $\disc{\overline{\mathbf{U}}}$ and $\disc{\bm{b}}$,
  respectively, with the brackets inside the norm omitted for clarity. Such a brief notation does not cause confusion, because norms in this work are always taken for the \emph{discretized vectors or matrices} with dimensions of $n\times 1$ or $n\times n$, respectively, and never for the velocity or force vectors at any particular location.}:
\begin{linenomath}
\begin{align}
& \frac{\| \delta \mathbf{U}\|}{\| \mathbf{U}\|}  
\le \mathcal{K}_\mathsf{A}  \frac{\|\delta \bm{b} \|}{\| \bm{b} \|}
\label{eq:global-cn}
\end{align}
\end{linenomath}
where
\begin{linenomath}
 \[
\mathcal{K}_\mathsf{A} \equiv \|\mathsf{A}\| \|\mathsf{A}^{-1}\|
\]
\end{linenomath}
denotes the condition number of matrix $\mathsf{A}$~\citep[see, e.g.,][]{strang93introduction}. 
Considering that the objective is to assess the effects of \emph{Reynolds stress perturbation} $\delta \boldsymbol{\tau}$ on the
velocities, the inequality in Eq.~(\ref{eq:global-cn}) above is formulated as follows:
\begin{linenomath}
\begin{equation}
  \label{eq:cond-norm-derive}
 \frac{\|\delta \mathbf{U}\|}{\|\mathbf{U}\|}  \le 
\mathcal{K}_\tau 
\frac{\|\nabla \cdot  \delta \boldsymbol{\tau} \|}{\| \nabla \cdot  \boldsymbol{\tau} \|} .
\end{equation}
\end{linenomath}
where
\begin{linenomath}
\begin{equation}
  \label{eq:cond-norm}
\mathcal{K}_\tau = \mathcal{K}_\mathsf{A} \frac{\|\nabla \cdot
  \boldsymbol{\tau}\|}{\|\bm{b}\|}
\end{equation} 
\end{linenomath}
and detailed derivations are omitted here for brevity and are presented in Appendix~\ref{app:global-cn}. It can be seen that the model condition number $\mathcal{K}_\tau$ consists of the condition number of the matrix $\mathsf{A}$
and the ratio in Eq.~(\ref{eq:alpha}). For plane channel flows the convective term disappears, and thus $\bm{b}$ is the force due to the divergence of the viscous stress, i.e., $\bm{b} = \nabla \cdot \nu (\nabla \bm{u}+(\nabla \bm{u})^T) =
\nabla \cdot \boldsymbol{\tau}_{\text{vis}}$. Consequently, the ratio $\overline{\alpha}$ indicates the overall relative importance of the forces due to Reynolds stress and viscous stress.
\begin{linenomath}
\begin{equation}
  \label{eq:alpha}
\overline{\alpha} = \|\nabla \cdot \boldsymbol{\tau}\| / \|\bm{b}\|.
\end{equation} 
\end{linenomath}

\begin{figure}
  \centering
  \hspace{1em}\includegraphics[width=0.35\textwidth]{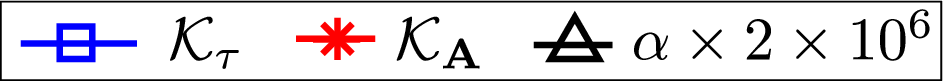}\\
  \includegraphics[width=0.48\textwidth]{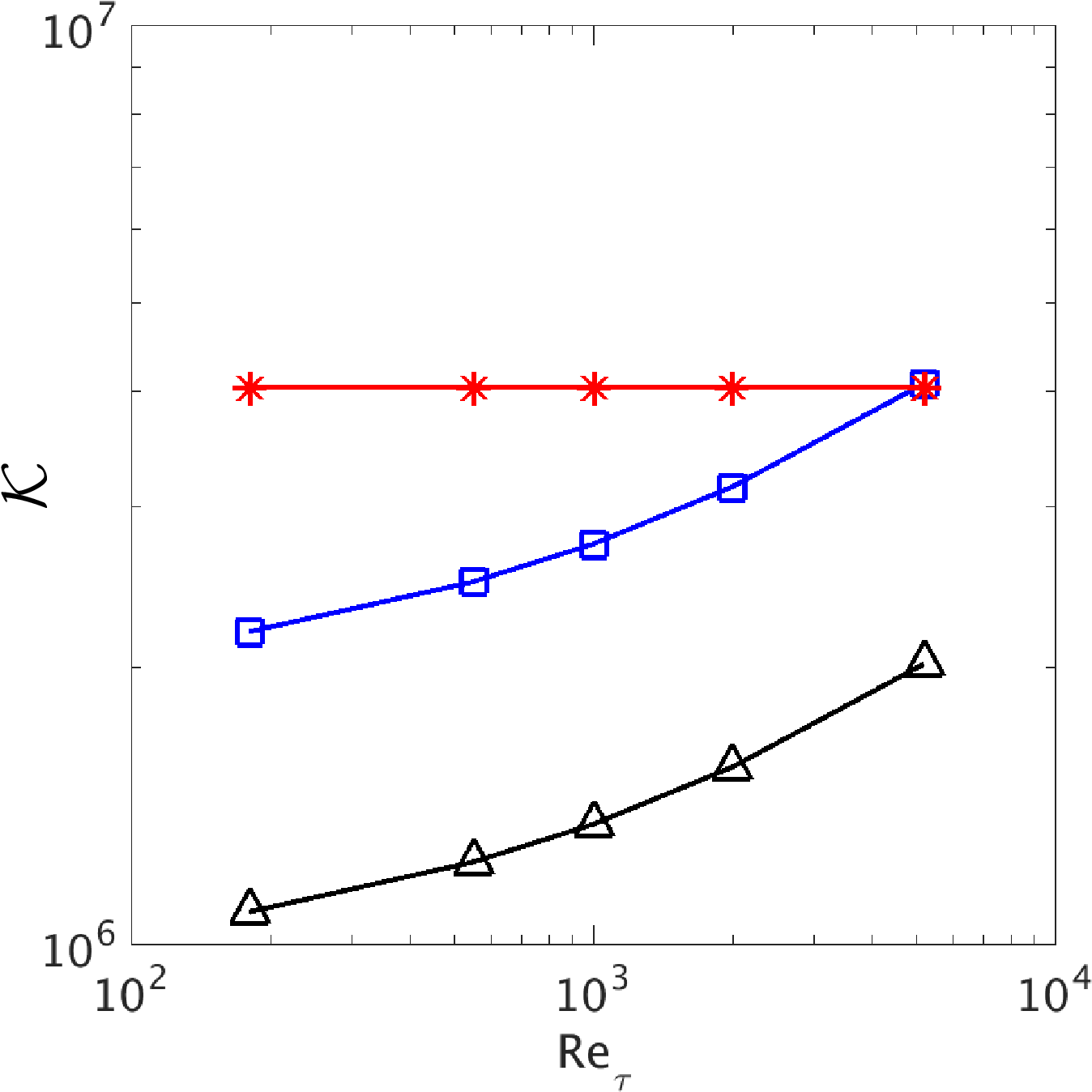}
    \caption{Conditioning measure of  Reynolds-stress-based turbulence models based on $\mathcal{K}_\mathsf{A}$ and the ratio $\overline{\alpha}$ as defined in Eq.~(\ref{eq:alpha}). The Reynolds stress is computed from DNS data to study the ideal scenario of the RANS modeling.}
  \label{fig:kappa-tau}
\end{figure}

The proposed condition number $\mathcal{K}_\tau$ based on matrix condition number $\mathcal{K}_\mathsf{A}$ is a natural first attempt in explaining the increasing sensitivity of the velocities to the Reynolds stress with increasing Reynolds numbers as shown in Table~\ref{tab:summary}. However, surprisingly it turns out that the condition number $\mathcal{K}_\tau$ is more or less the same across all Reynolds numbers from $Re_\tau = 180$ to $5200$, which is shown in Fig.~\ref{fig:kappa-tau}. This observation suggests that the matrix-based condition number $\mathcal{K}_\tau$ cannot explain the ill-conditioning of the $Re_\tau = 5200$ case and the better conditioning of the lower Reynolds number cases as observed in Table.~\ref{tab:summary}.

The following two factors explain why the matrix-based condition number $\mathcal{K}_\tau$ is almost the same at different Reynolds numbers:
\begin{enumerate}[(1)]
\item the matrix condition number $\mathcal{K}_\mathsf{A}$ is constant for all Reynolds numbers, because the matrix $\mathsf{A}$ itself is independent of the Reynolds number.
\item The ratios \(\overline{\alpha} = \frac{\|\nabla \cdot
  \boldsymbol{\tau}\|}{\|\bm{b}\|}\) are very similar at vastly different Reynolds numbers, because both norms (which involve integration or square sums) are dominated by the \emph{viscous wall regions} of each flow. It is well-known that the Reynolds number only determines the thickness of this region in outer coordinates, and the Reynolds number effect is weak here. Each factor above will be detailed as follows.
\end{enumerate}

First, it can be established through simple algebra that for plane channel flows the matrix $\mathsf{A}$ is independent of the Reynolds number but depends on the discretization scheme and the mesh used. Since the convection term disappears for plane channel flows, the matrix $\mathsf{A}$ results solely from the discretization of the diffusion operator $\nu \nabla^2 (\cdot)$. When discretized with central difference on a uniform mesh of $n$ cells, matrix $\mathsf{A}$ can be written as follows~\citep{strang93introduction}:
\begin{linenomath}
\begin{equation}
\mathsf{A} = \nu
\begin{bmatrix}
\begin{array}{ccccc}
  2  &  -1  &       &       &    \\ 
 -1  &   2  & -1    &       &    \\ 
     &  -1  &  2    &  \ddots &     \\
     &      & \ddots & \ddots & -1  \\
     &      &    &    -1 & 2
\end{array}
\end{bmatrix}.
\label{eq:simple-matrix}
\end{equation}
\end{linenomath}
The condition number for matrix $\mathsf{A}$ is $\mathcal{K}_\mathsf{A} = 4 n^2 / \pi^2$. Therefore, $\mathcal{K}_\mathsf{A}$ does not explicitly depend on the viscosity or the Reynolds number. This analysis is confirmed by the results shown in Fig.~\ref{fig:kappa-tau}, which shows that $\mathcal{K}_\mathsf{A}$ is strictly constant for all five flows at different Reynolds numbers. Moreover, $\mathcal{K}_\mathsf{A}$ depends on the mesh size $n$, which is a critical shortcoming of the matrix-norm based condition number as a measure of the conditioning property of a turbulence model. To exclude the influences of the mesh, we used the same mesh with 1040 cells in all the flows at different Reynolds numbers presented in Fig.~\ref{fig:kappa-tau}.

\begin{figure}
  \centering
  \includegraphics[width=0.5\textwidth]{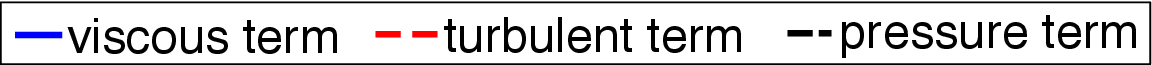}\\
  \subfloat[$Re_\tau=180$]{\includegraphics[width=0.45\textwidth]{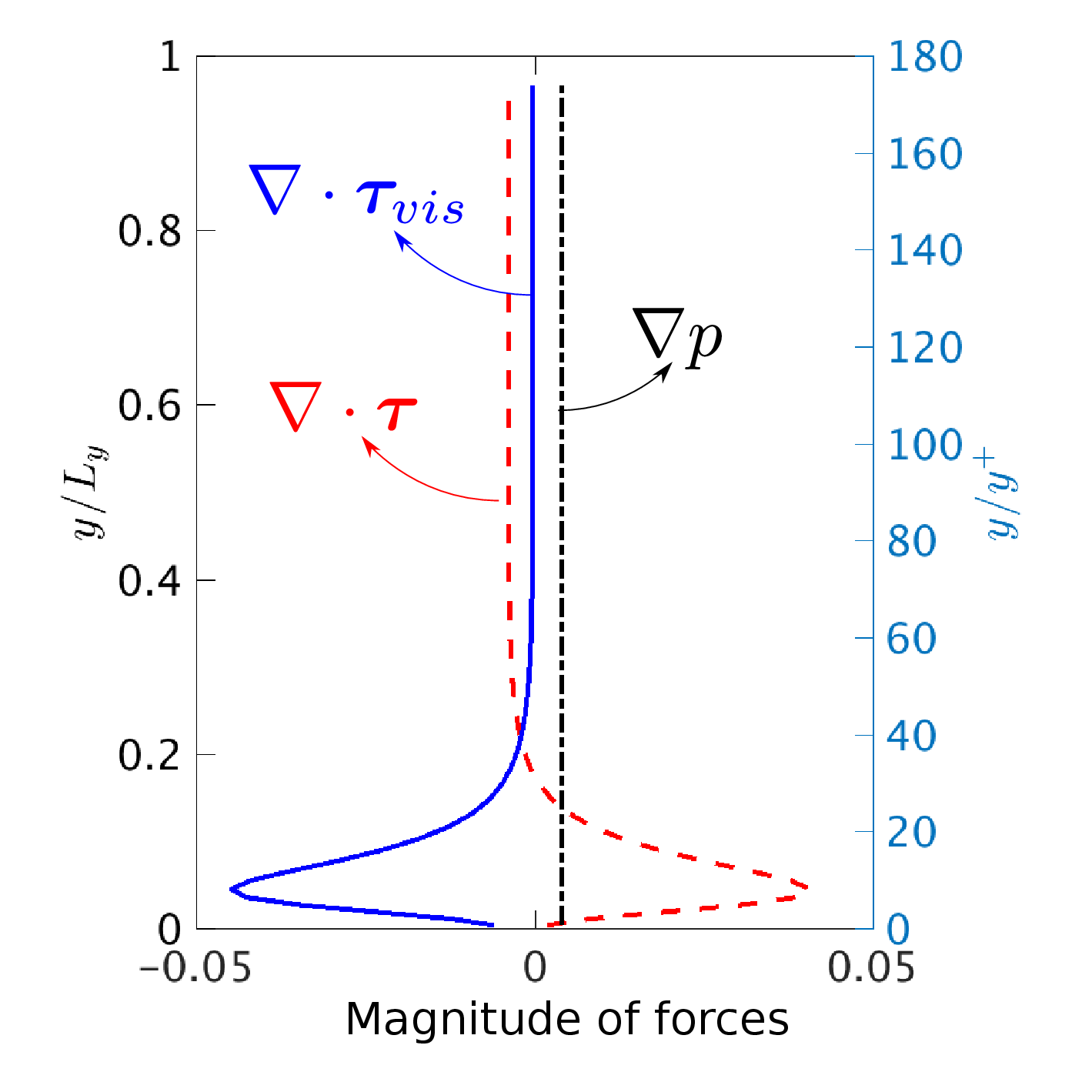}}\hspace{1em}
  \subfloat[$Re_\tau=5200$ (bottom 1/10 channel)]{\includegraphics[width=0.45\textwidth]{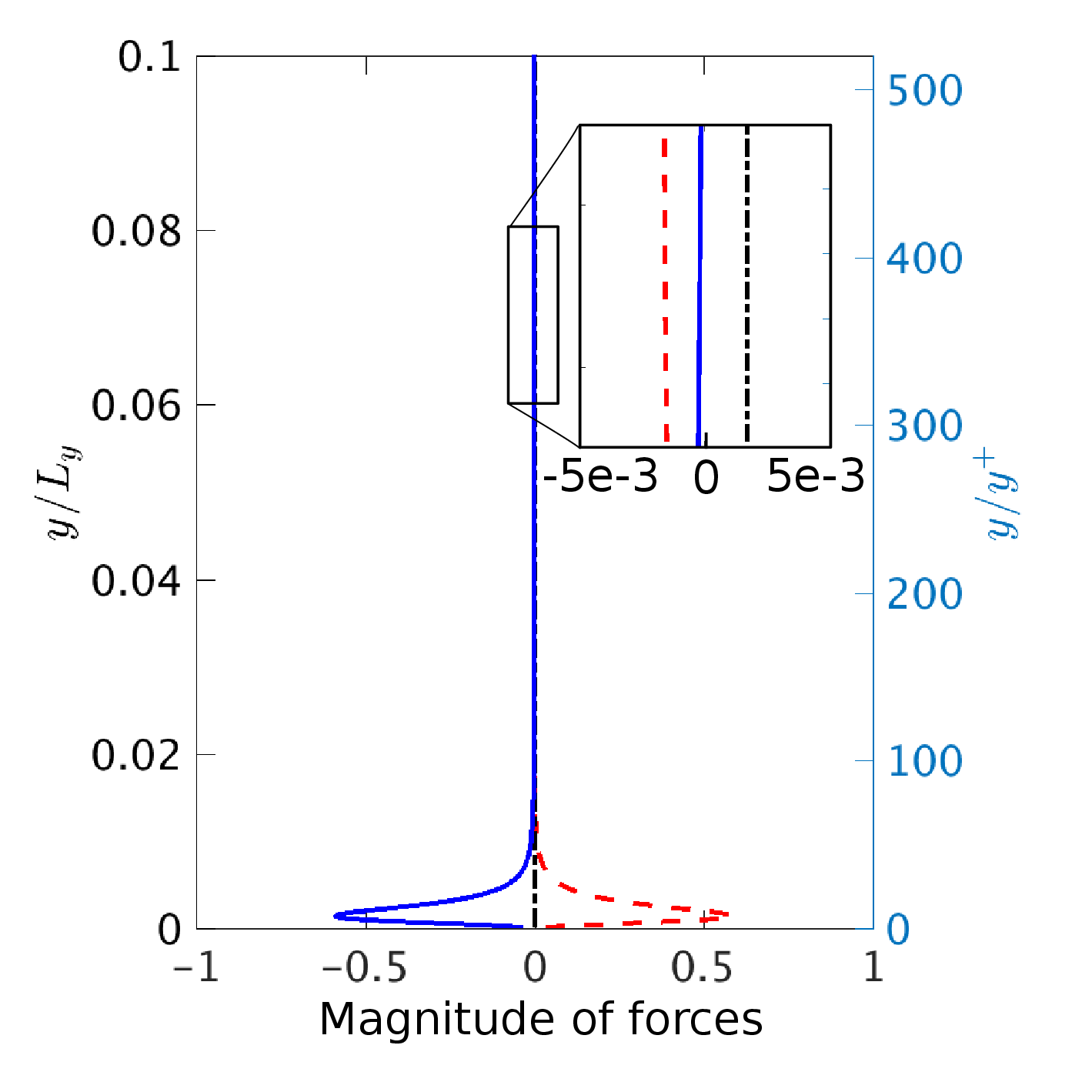}}
    \caption{The balance among forces due to turbulent shear stress ($\nabla \cdot \bm{\tau}$), viscous shear stress ($\nabla \cdot \bm{\tau}_\text{vis}$), and pressure gradient ($\nabla p$) for two plane channel flows at frictional Reynolds numbers (a) $Re_\tau=180$ and (b) $Re_\tau=5200$. The right vertical axis denotes the inner coordinates ($y^+$).}
  \label{fig:force-profiles}
\end{figure}

\begin{figure}
  \centering
      \includegraphics[width=0.95\textwidth]{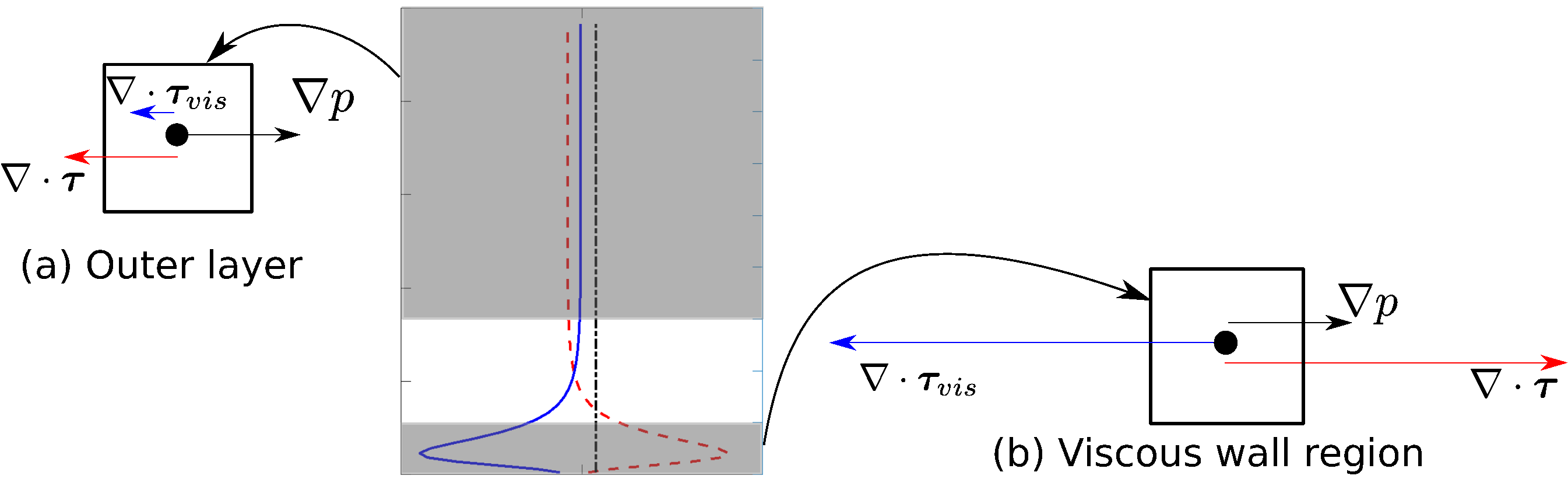}
    \caption{Force balance of the plane channel flow in (a) the outer layer and (b) the viscous wall region.}
  \label{fig:balance}
\end{figure}

Second, the change of Reynolds number has little influence on the ratio $\overline{\alpha}$, as is shown in Fig.~\ref{fig:kappa-tau}. We examine the profile of turbulent shear ($\nabla \cdot \boldsymbol{\tau}$) and viscous shear ($\nabla \cdot \boldsymbol{\tau}_\text{vis}$) in the channel in Fig.~\ref{fig:force-profiles} for the two cases, $Re_\tau = 180$ and 5200. In most of the channel outside the viscous wall region, both forces (and thus the ratio) are fairly uniform. Nevertheless, in the viscous wall region, the two forces are of the same order of magnitude but with opposite signs. In contrast, outside the viscous region, the pressure gradient is the main driving force while the Reynolds shear stress is the resistance, with the viscous shear having negligible effects. The forces in the two distinct regions are illustrated schematically in Fig.~\ref{fig:balance}.  However, when calculating the ratio $\overline{\alpha} =  \|\nabla \cdot \boldsymbol{\tau}\|/\|\bm{b}\|$ of the two norms, values within the viscous wall region clearly dominates the calculation of both norms, which involve integration of the functions squared. It can be seen that in both Fig.~\ref{fig:force-profiles}a  ($Re_\tau = 180$) and Fig.~\ref{fig:force-profiles}b ($Re_\tau=5200$) the areas enclosed by the blue/solid curve and red/dashed curve (with the vertical zero line) are similar. Squaring the function places even more weights on the regions of larger function values, i.e., the viscous wall region. This observation suggests that the ratio $\|\nabla \cdot \boldsymbol{\tau}\|/\|\bm{b}\|$ is of order $O(1)$ for both cases, as confirmed by examining Fig.~\ref{fig:kappa-tau}. Consequently, the computed norm mostly reflects the values of the forces in the viscous region and not the outer layer.  It is well known that the Reynolds number effects are not pronounced within the viscous wall region. Increasing the Reynolds number merely extends the outer layer in terms of inner coordinates ($y^+ = y/y^*$ where $y^*=\nu/\sqrt{\tau_w/\rho}$ is the viscous unit and $\tau_w$ is the wall shear stress). This explains why the ratio $\overline{\alpha}$ does not vary significantly (much less than proportionally) with the Reynolds number as can be seen in Fig.~\ref{fig:kappa-tau}. If the viscous region is neglected, the factor $\bar{\alpha}$ will be higher for the larger Reynolds number, and thus it makes the global condition number a better indicator of model conditioning. However, the matrix condition number $\mathcal{K}_\mathsf{A}$ depends on the mesh size and thus the global condition number is still not an ideal choice of evaluating model conditioning of RANS equations.

In summary, the matrix-based condition number $\mathcal{K}_\tau$ as derived in Eq.~(\ref{eq:cond-norm}) is not able to explain the increasing sensitivity of the velocities with respect to Reynolds stresses with increasing Reynolds number. In addition, the matrix condition number $\mathcal{K}_\mathsf{A}$ has another critical drawback of being mesh dependent. The mesh dependency is highly undesirable as the condition number is to measure the conditioning property of \emph{turbulence models} at the PDE level, not any particular numerical discretization thereof.
These drawbacks clearly call for a better metric for measuring the conditioning property of Reynolds-stress-based turbulence models. 

\subsection{Proposed metric as a measure of model conditioning}
\label{sec:local-cn}
In order to address the deficiency of the global condition number as presented in Section~\ref{sec:global-cn}, we derive a metric based on a local condition number function to measure the sensitivity of the solved mean velocity  $\bm{u}$  at a given location $\bm{x}$ with respect to perturbation $\delta \ms{\tau}$ on the Reynolds stresses field $\bm{\tau}$. Such a local condition number is formally defined as the following bound:
\begin{linenomath}
\begin{align}
\frac{|\delta \bm{u}(\mb{x})|}{U_\infty} 
& \le 
\mathcal{K}(\mb{x}) \,
\frac{\Cnorm{\nabla \cdot \delta \ms{\tau}}}
{\Cnorm{\nabla \cdot \ms{\tau}}}
\label{eq:cond-derive}
\end{align}
\end{linenomath}
where $\mathcal{K}(\mb{x})$ is the local condition number function defined as:
\begin{linenomath}
\begin{equation}
\label{eq:cont-cond}
\mathcal{K}(\mb{x}) = 
\frac{\Cnorm{G({\mb{x}}, \bm{\xi})} \,
\Cnorm{\nabla \cdot \ms{\tau}}}{U_\infty}
\end{equation}
\end{linenomath}
Function $G$ is the Green's function corresponding to the linear operator $\mathcal{L}$~\citep[see, e.g.,][]{lanczos96linear}, such that the
solution to the linearized RANS equation~(\ref{eq:ns-concise-L}) can be written formally as:
\begin{linenomath}
\begin{equation}
\label{eq:define-green}
\bm{u}(\mb{x})  = \mathcal{L}^{-1} [\mb{b}(\mb{x})]
 \equiv \int_\Omega G(\mb{x}; \ms{\xi}) \, \mb{b}(\ms{\xi}) \, d\ms{\xi}
\end{equation}
\end{linenomath}
where $\mathcal{L}^{-1} [\cdot]$ is the inverse operator of $\mathcal{L}$. Green's function $G(\mb{x}; \ms{\xi})$ indicates the contribution of the source $\mb{b}(\ms{\xi})$ at location $\ms{\xi}$ to the solution $\bm{u}$ at location $\mb{x}$. The norm $\Cnorm{f(\ms{\xi})}$ of function $f(\ms{\xi})$ is an integration on domain $\Omega$ defined as~\citep{debnath05hilbert}:
\begin{linenomath}
\begin{equation}
\label{eq:function-norm-def}
\Cnorm{f(\ms{\xi})}=\sqrt{\int_\Omega |f(\ms{\xi})|^2 d \ms{\xi}} \; .   \end{equation}
\end{linenomath}
The detailed derivations to obtain Eq.~(\ref{eq:cont-cond}) are presented in Appendix~\ref{sec:app-local-cn}.

For functions discretized on a CFD mesh of $n$ cells (e.g., those in RANS simulations), the function norm $\| \cdot \|_\Omega$ in Eq.~(\ref{eq:cont-cond}) can be approximated by the norm of the discretized $n$-vector through numerical quadrature. That is,
\begin{linenomath}
\begin{equation}
\Cnorm{G(\bm{x}, \bm{\xi})} \approx \Vnorm{{\mb{r}}_j} 
\qquad  
\text{and} 
\qquad 
\Cnorm{\nabla \cdot \ms{\tau}} \approx \Vnorm{[\nabla \cdot \ms{\tau}]}
\end{equation}
\end{linenomath}
where $\mb{r}_{j}$ is the $j$-th row of the matrix $\mathsf{A}^{-1}$. Recall that $[\nabla \cdot \ms{\tau}]$ indicates discretization of field $\nabla \cdot \ms{\tau}$ on the CFD mesh, but the bracket can be omitted inside a vector norm $\Vnorm{\cdot}$ without ambiguity. In this work, the norm $\|\cdot\|_n$ is defined by $L^2$ norm as follows:
\begin{linenomath}
\[
\|\mathbf{v}\|_n=\left(\sum_i v_i^2\right)^{1/2}
\]
\end{linenomath}
and the discretized condition number $n$-vector corresponding to $\mathcal{K}(\bm{x})$ in Eq.~(\ref{eq:cont-cond}) is thus:
\begin{linenomath}
\begin{equation}
\label{eq:cont-cond-disc}
\mathcal{K}_j =  
\frac{\Vnorm{\mb{r}_{j}} \,
\Vnorm{\nabla \cdot \ms{\tau}}}{U_\infty}
\qquad \text{with} \quad j = 1, 2, \cdots, n
\end{equation}
\end{linenomath}
which implies that the location $\mb{x}$ is the coordinate of the $j$-th cell in the CFD mesh. 

The proposed local condition number function $\mathcal{K}(\bm{x})$ has two important merits compared to the global matrix based condition number $\mathcal{K}_\tau$: 
\begin{enumerate}[(1)]
    \item $\mathcal{K}(\bm{x})$ provides a tighter upper bound than the matrix-norm condition number $\mathcal{K}_{\tau}$. The main reason is that the upper bound of $\mathcal{K}_{\tau}$ can only be achieved when the following conditions are satisfied simultaneously: (i) the discretized mean velocity field vector $\mathbf{U}$ is aligned with the principal axis of the coefficient matrix $\mathsf{A}$, and (ii) that the perturbation vector $\delta \bm{b}$ is aligned with the principal axis of $\mathsf{A}^{-1}$. 
    In contrast, the derivation of $\mathcal{K}(\bm{x})$ does not assume any conditions on the discretized mean velocity field $\mathbf{U}$. Consequently, the bound provided by $\mathcal{K}(\bm{x})$ is a more precise assessment of the sensitivity $\delta \bm{u}$ with respect to Reynolds stress perturbations.
  \item The discretization $\mathcal{K}_j$ of function $\mathcal{K}(\bm{x})$ is mesh indepedent, which is an important property considering that this metric aims to measure the conditioning property of data-driven Reynolds stress models. Detailed analytical derivations to obtain Eq.~(\ref{eq:cont-cond-disc}) and numerical results (see Fig.~\ref{fig:mesh-converge} in the appendix) used to demonstrate the mesh independency of the local condition number $K_j$ are presented in Appendix~\ref{sec:discretization}.
\end{enumerate}

While the local condition metrics proposed here are generally applicable to any linearized PDE and its discretization, it is important to emphasize that these conditioning metrics faithfully embody the \emph{physics} described by the underlying PDEs.  Taking the RANS equations as an example, the mean flow field contributes to the linearized differential operator and thus is reflected in the analysis of local conditioning. Therefore, these local condition metrics are flow-specific properties and thus reflect the mean flow physics. More detailed discussion can be found in Sections~\ref{sec:res-duct} and \ref{sec:res-pehill} on the complex, two-dimensional flows. 

Finally, we comment briefly regarding the numerical implementation and computational complexity of conditioning metrics proposed below. The local condition number $\mathcal{K}_j$ does not require a full inversion of the matrix $\mathsf{A}$ but only needs the $j$-th row of $\mathsf{A}^{-1}$ to be computed. This can be achieved by solving the equation $\mathsf{A}^\textrm{T}\mathbf{r}_j=\mathsf{I}_j$ based on the identity $\mathsf{A}^{-1}\mathsf{A}=\mathsf{I}$, where $\mathsf{I}_j$ denotes the $j$-th column of the identity matrix $\mathsf{I}$. Solving this equation is a standard, inexpensive routine available in CFD codes, e.g., with a computational complexity of $\mathcal{O}(n \log n)$ if a multigrid linear solver is used, where $n$ indicates the number of elements in $\mathbf{r}$, and $\mathcal{O}$ denotes ``of the order of''. Therefore, it can be estimated that even obtaining the full condition number field only has a computational complexity of $\mathcal{O}(n^2 \log n)$ with a standard multigrid linear solver, which is much lower than the complexity of $\mathcal{O}(n^3)$  for typical algorithms of matrix inversions.

As the local condition number $\mathcal{K}(\bm{x})$ is a spatial function, and its discretization is an $n$-vector, it is desirable to obtain a scalar quantity to provide an integral, more straightforward measure of model conditioning property similar to the global condition number $\mathcal{K}_{\tau}$. To this end, we define a
\emph{volume-averaged} local condition number $\overline{\mathcal{K}}_{\bm{x}}$ defined in Eq.~(\ref{eq:avg-kappa}).
\begin{linenomath}
\begin{equation}
\overline{\mathcal{K}}_{\bm{x}} = 
\frac{\sum_{j=1}^{n}[\mathcal{K}_j] \, [\Delta V_j]}{V}
\label{eq:avg-kappa}
\end{equation}
\end{linenomath}
where $\Delta V_j$ denotes the volume of the $j$-th cell in the CFD mesh, and $V$ is the total volume of the computational domain. This volume-averaged local condition number $\overline{\mathcal{K}}_{\bm{x}}$ has a similar interpretation to $\mathcal{K}_{\tau}$ but preserves the merits of $\mathcal{K}_j$, i.e., tighter bounds and mesh independency.

In the derivations above the Reynolds stress term is substituted directly into the RANS equation and is treated explicitly. When the Reynolds stress term is treated implicitly as in many practical implementations of Reynolds stress models~\citep[e.g.,][]{basara03new,maduta17improved}, the corresponding local condition number of the model can be obtained similarly, except that the Green's function is modified to account for the implicit modeling of the linear part of Reynolds stress with eddy viscosity model. Specifically, the general form of implicit treatment of Reynolds stress can be written as follows:
\begin{linenomath}
\begin{equation}
\label{eq:nut-model}
\ms{\tau} = 2\nu_t \mathbf{S} + \ms{\tau}^{\perp}
\end{equation}
\end{linenomath}
where $\nu_t$ represents the eddy viscosity, $\mathbf{S}=\frac{1}{2}\left( \nabla \bm{u} + (\nabla \bm{u})^T\right)$ denotes the strain rate tensor and $\ms{\tau}^{\perp}$ denotes the nonlinear part. In this work, the eddy viscosity $\nu_t$ is obtained from projecting DNS Reynolds stress onto DNS strain rate tensor and not from RANS simulations. With such an optimal eddy viscosity $\nu_t^m$, Eq.~(\ref{eq:nut-model}) treats the linear part of Reynolds stress tensor implicitly to enhance the conditioning of RANS equations. Consequently, the Green's function
$\widetilde{G}$ corresponding to the linear operator
\begin{linenomath}
\begin{equation}
\label{eq:L-implicit}
\widetilde{\mathcal{L}}(\bm{u}) =
\mathcal{L}(\bm{u}) - \nu_t^m \nabla^2 \bm{u}
=
\bm{u}_0 \cdot  \nabla \bm{u} - (\nu + \nu_t^m) \nabla^2 \bm{u} 
\end{equation}
\end{linenomath}
should be used in Eqs.~(\ref{eq:cont-cond}) and (\ref{eq:cont-cond-disc}), with $\widetilde{G}$ related to $\widetilde{\mathcal{L}}$ in a similar way as $G$ to $\mathcal{L}$ in Eq.~(\ref{eq:define-green}). The optimal eddy viscosity $\nu_t^m$ is computed by minimizing the discrepancy between the linear eddy viscosity model and the DNS Reynolds stress data, i.e.,
\begin{linenomath}
\begin{equation}
\label{eq:nut-definition}
\nu_t^m(\mb{x}) = \argmin_{\nu_t} || \ms{\tau}^\text{DNS} - 2\nu_t(\mb{x}) \mathbf{S}^\text{DNS}||
\end{equation}
\end{linenomath}
where $\bm{\tau}^\text{DNS}$ and $\mathbf{S}^\text{DNS}$ denote the Reynolds stress and the strain rate tensor from DNS database, respectively. Here we emphasize that the optimal eddy viscosity is location-dependent. The detailed derivations are presented in Appendix~\ref{sec:app-eddy-cn}. Noted that the eddy viscosity $\nu_t$ in Eq.~(\ref{eq:nut-model}) only quantifies the amount of Reynolds stress being treated implicitly and is not necessarily the optimal eddy viscosity $\nu_t^m$. By specifying different $\nu_t$, the amount of implicit treatment of Reynolds stress and the amount of nonlinear part of Reynolds stress vary accordingly, leading to different conditioning of RANS equations. Therefore, Eq.~(\ref{eq:nut-model}) provides a general form of implicit treatment of Reynolds stresses. The optimal eddy viscosity is capped to be positive for numerical stability of solving the RANS equations. Therefore, the implicit treatment always leads to better conditioning of the system, but the difference between implicit treatment and explicit treatment becomes smaller in the regions where linear part of Reynolds stress is less dominant. In the extreme (albeit unlikely) situation where Reynolds stress tensor is orthogonal to the strain rate tensor (i.e., optimal eddy viscosity is zero across the flow domain), the conditioning with implicit treatment and explicit treatment would be equivalent.

In summary, we proposed a local condition number to assess the sensitivity of local mean velocities with regard to data-driven Reynolds stress models. It has the following three forms: the spatial function $\mathcal{K}(\bm{x})$ (i.e., condition number function), an $n$-vector $\mathcal{K}_j$ obtained by discretizing $\mathcal{K}(\bm{x})$ on the CFD mesh, and a scalar $\overline{\mathcal{K}}_{\bm{x}}$ obtained by integration of $\mathcal{K}(\bm{x})$. This metric is applicable to different types of data-driven Reynolds stress models. The main contributions of this work are (1) highlighting the importance of the conditioning of PDE-governed systems with data-driven closure models and (2) providing a quantitative metric for assessing such conditioning. Data-driven modeling is becoming an emerging area in the fluid mechanics community. However, all existing works of data-driven modeling focus on the accuracy of the model itself.  Our work demonstrates that an accurate data-driven model does not necessarily guarantee satisfactory predictions of quantities of interest.  Therefore, ensuring good conditioning of the problem formulation (i.e., PDEs with data-driven closure) is as important as improving the accuracy of data-driven model. We envision a standard practice in the future where all developed data-driven closures are presented along with corresponding conditioning analysis, in the same way in which experimental data reported nowadays are accompanied by their associated uncertainties.

\section{Results}
\label{sec:results}
We first use the proposed local condition number to explain the model conditioning of RANS equations when using explicit treatment with fixed Reynolds stress. Specifically, turbulent channel flows at different Reynolds numbers are studied and the results are shown in Table~\ref{tab:summary}. In addition, the implicit treatment of Reynolds stress is studied to compare with the model conditioning of RANS equations by using explicit treatment with fixed Reynolds stress. By studying these two types of RANS modeling, we show that the proposed local condition number can be used to assess the sensitivities of RANS simulations for data-driven modeling. We also extend the study of the model conditioning and the proposed local condition number to two more complex flows, including the flow over periodic hills and the flow in a square duct. The results show that the RANS equations can be ill-conditioned for other types of flows, which can be assessed by the proposed condition number metric.

The RANS simulations are performed in a finite-volume CFD platform OpenFOAM, using a modified flow solver that allows the explicit and implicit treatments of Reynolds stress computed from DNS data. For numerical discretizations, the second-order central difference scheme is chosen for all terms except for the convection term, which is discretized with a second-order upwind scheme. The second-order upwind scheme was used to avoid the possible numerical instability when using central difference scheme for the convection term. For the turbulent plane channel cases, the convection term disappears at steady state and thus the presented results in Section~\ref{sec:data-driven} are not influenced by the discretization of the convection term thereof. In Section~\ref{sec:data-driven}, the mean velocity is obtained by directly solving Eq.~\ref{eq:ns-concise-L} since the mean velocity and the pressure are decoupled for the RANS simulation of a fully-developed plane channel flow. The convergence criteria of solving the momentum equations is set as~$10^{-8}$ in absolute error.

\subsection{Turbulent channel flow}
\label{sec:data-driven}
The fully developed turbulent plane channel flows are investigated by using the local condition number $\mathcal{K}_j$. In this work, we consider an ideal scenario in which the Reynolds stress $\bm{\tau}$ is directly computed from DNS database at various Reynolds numbers $Re_{\tau}=180, 550, 1000, 2000, 5200$. The numbers of mesh cells $N$ are $36, 110, 200, 400, 1040$, respectively, for the channel flows at Reynolds numbers $Re_{{\tau}}=180, 550, 1000, 2000$, and $5200$. Non-uniform meshes are used and the expansion ratio is adjusted to ensure that the $y^+$ of the first cell center is kept below 1. Therefore, the mesh size of the first cell at the wall-normal direction is below $h/N$, where $h$ denotes the half channel height. The DNS data were obtained from the University of Texas Austin online database~\citep{lee15direct}. The mean velocity field is then solved by substituting the computed Reynolds stress as the closure term of RANS equations.

In the practice of RANS modeling, iterations are involved in solving RANS equations and the modeling of Reynolds stress is updated by the mean velocity field during the iterations. Therefore, it is possible that the mean velocity field and the Reynolds stress can adjust to each other during the iterations. We employed the ratio $\delta U_{rms}/U^\text{DNS}_{rms}$ to assess the error of the solved mean flow field at each iteration step. Specifically, the volume-averaged root-mean-squared error of the solved mean velocity is defined as follow:
\begin{linenomath}
\begin{equation}
\delta U_{rms} = 
\sqrt{\frac{\sum_{j=1}^{n} \left([U]_j-[U^\text{DNS}]_j\right)^2 \, [\Delta V_j]}{V}}
\label{eq:deltaUrms}
\end{equation}
\end{linenomath}
The volume-averaged root-mean-squared DNS velocity is defined as follow:
\begin{linenomath}
\begin{equation}
U^\text{DNS}_{rms} = 
\sqrt{\frac{\sum_{j=1}^{n}\left([U^\text{DNS}]_j \right)^2 \, [\Delta V_j]}{V}}
\label{eq:Urms}
\end{equation}
\end{linenomath}

\subsubsection{Reynolds stress models with explicit treatment}
\label{sec:explicit-treatment}
The Reynolds stress term is directly computed from DNS data and substituted into RANS as shown in Algorithm~\ref{alg:reynolds-stress-model}. The purpose is to study most existing data-driven RANS models, in which the Reynolds stress is directly predicted by training on DNS data and then used to solve for mean velocity field. In this part, we only study the explicit treatment with fixed Reynolds stress, i.e., the dependence of Reynolds stress on strain rate tensor is not considered. It is because such a dependence has not been taken into consideration in many data-driven turbulence models, and thus we illustrate the corresponding issue here. More results of explicit treatment with dependence on strain rate tensor can be found in Appendix~\ref{sec:app-explicit-dependence}, where we further show that the explicit coupling between Reynolds stress and mean velocity during the iterations can gradually amplify the errors.

\vspace{1em}
\begin{minipage}{0.95\textwidth}
\begin{algorithm}[H]
  Set Reynolds stress from DNS data: $\bm{\tau}=\bm{\tau}^\text{DNS}$ \\
  \For{\textrm{each iteration step} $i=1,2,...,N$}{
  Solve RANS equations: $\mathcal{N}\left(\overline{u}^{(i)}\right) = \nabla \cdot \bm{\tau}- \nabla p$ to obtain $\overline{u}^{(i)}$
  }
  \caption{\emph{Explicit} treatment of Reynolds stress with the DNS Reynolds stress being \emph{fixed} among iterations.}
  \label{alg:reynolds-stress-model}
\end{algorithm}
\end{minipage}
\vspace{1em}

\begin{figure}
  \centering
  \hspace{2em}\includegraphics[width=0.4\textwidth]{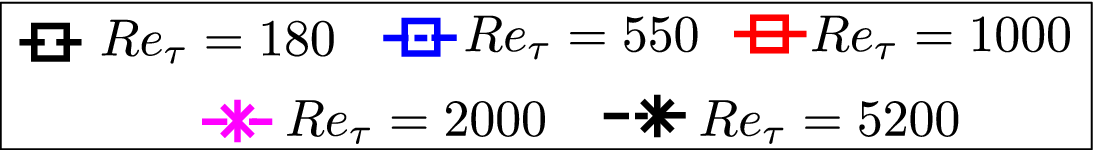}\\
  \includegraphics[width=0.45\textwidth]{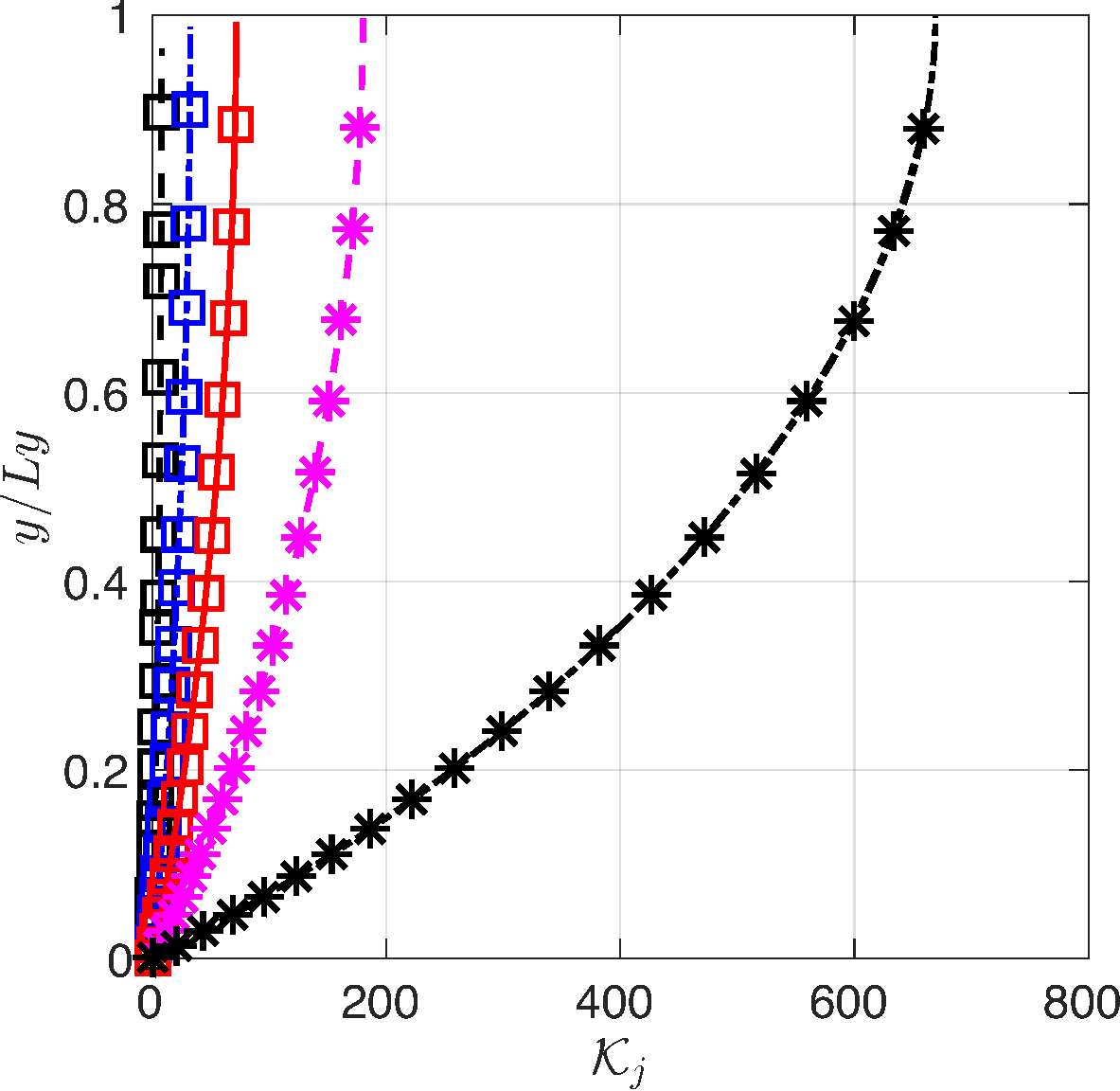}
    \caption{The profiles of local condition number $\mathcal{K}_j$ at different Reynolds numbers by using \textbf{explicit treatment} with fixed Reynolds stress, i.e., the fixed DNS Reynolds stress is substituted into RANS equations.}
  \label{fig:cmptKappaTau}
\end{figure}

The local condition numbers $\mathcal{K}_j$ of the explicit treatment of Reynolds stress are shown in Fig.~\ref{fig:cmptKappaTau}. With the increase of the Reynolds number, it can be seen that the magnitude of local condition number also increases. Specifically, the local condition number of the flow at $Re_\tau=180$ is of the order $O(1)$, while the local condition number of the flow at $Re_\tau=5200$ is of the order $O(10^2)$. This rapid increase of local condition number agrees well with the increased error of solved mean velocity $U$ as summarized in Table~\ref{tab:summary}. In addition, the local condition number is greater near the channel center than close to the wall, especially for the high Reynolds number cases. Such pattern of local condition number also agrees with the spatial pattern of the error of the solved mean velocity as illustrated in Fig.~\ref{fig:U-comp-nut}b. As demonstrated by~\cite{thompson16methodology}, the error of solved channel flow mean velocity at a given point is an integration of Reynolds stress errors (which happen to be of the same sign in the entire domain, indicating a systematic nature) from the wall to that given point, i.e., the Reynolds stress errors in between are \emph{accumulated} without cancellation. The local condition number in Fig.~\ref{fig:cmptKappaTau} faithfully reflects such an accumulative nature with deteriorated conditioning further away from the wall for all cases.  Recall that the local condition number $\mathcal{K}_j$ assesses the relative error of solved mean velocity at a given point with respect to the error in the whole Reynolds stress field, and not with respect to the error of Reynolds stress at the same given point.

\begin{figure}
  \centering
  \subfloat[Relative error]{\includegraphics[height=0.35\textwidth]{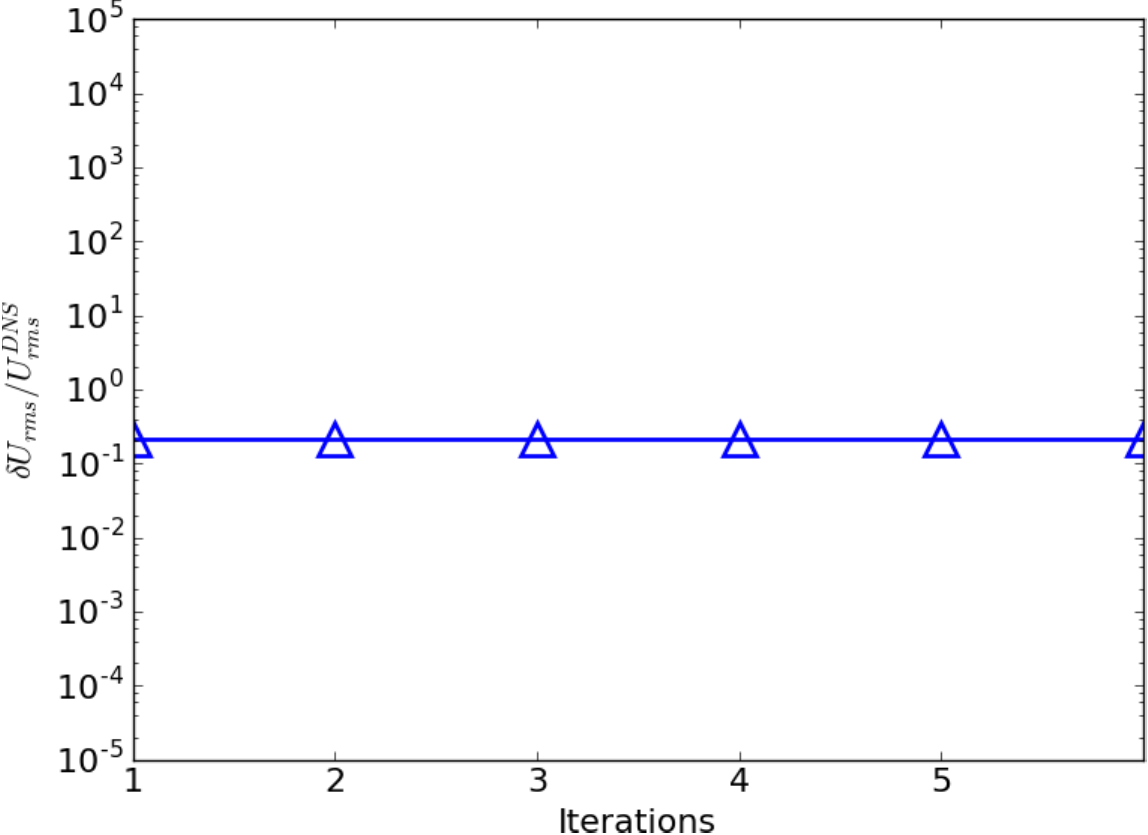}}\hspace{0.5em}
  \subfloat[Condition number]{\includegraphics[height=0.35\textwidth]{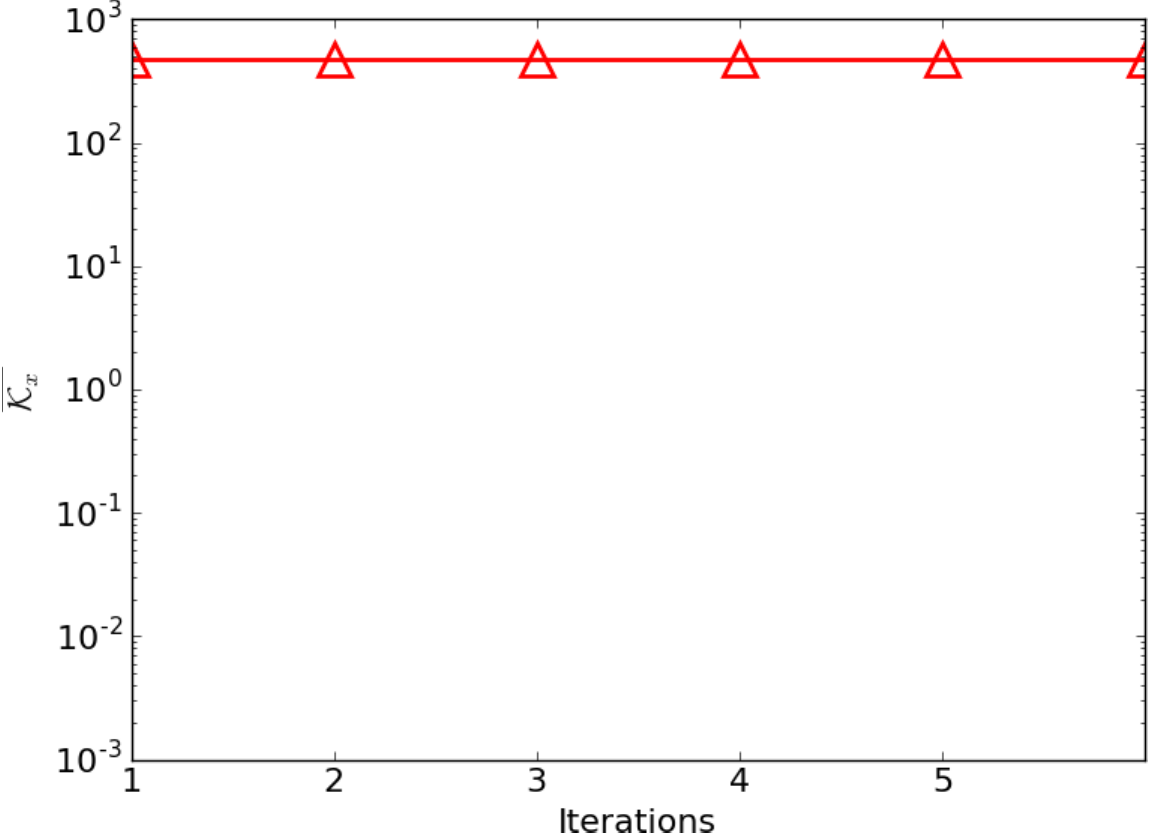}}
    \caption{The error propagation analysis of solving stream-wise velocity iteratively by using \textbf{explicit treatment} with fixed Reynolds stress, including (a) relative error of mean velocity and (b) volume-averaged local condition number.}
  \label{fig:frozen-Ux}
\end{figure}

The averaged local condition number $\overline{\mathcal{K}}_{\bm{x}}$ increases with the Reynolds number by using explicit treatment of Reynolds stress, wich is clearly seen in Fig.~\ref{fig:avg-cmptKappaTau-nut}. Such increase of averaged local condition number with Reynolds number reveals the potential shortcoming of explicit modeling of Reynolds stress, i.e., a relatively accurate but explicit modeling of Reynolds stress does not guarantee the satisfactory mean velocity by solving RANS equations, especially for high Reynolds number flows. This observation has been reported in the work of \cite{thompson16methodology}, and the proposed averaged local condition number can be used as an integral indicator to estimate the extent of error propagation from the modeled Reynolds stress to the solved mean velocity field. The error propagation with iterations is presented in Fig.~\ref{fig:frozen-Ux}a together with the averaged local condition number in each iteration. It can be seen that the error in the solved mean velocity stays constant in every iteration step. This observation is consistent with our expectation since the convection term disappears in the channel flows and the RANS equations become linear. Therefore, the error within the solved mean velocity in a given iteration does not influence the model conditioning and thus has no effect upon the error in the solved mean velocity at the next iteration.

\subsubsection{Reynolds stress models with implicit treatment}
\label{sec:evm}
The eddy viscosity is directly computed from DNS data and substituted into RANS equations to study the ideal situation of data-driven Reynolds stress models with implicit treatment as shown in Algorithm~\ref{alg:eddy-viscosity-model}.

\vspace{1em}
\begin{minipage}{0.95\textwidth}
\begin{algorithm}[H]
Compute optimal eddy viscosity $\nu_t^m$ from DNS Reynolds stresses based on Eq.~(\ref{eq:nut-definition}) \\
\For{\textrm{each iteration step} $i=1,2,...,N$}{
Compute Reynolds stress: $\bm{\tau}^{(i)} = \nu_t^m \left(\nabla \overline{u}^{(i)}+(\nabla \overline{u}^{(i)})^T\right) + \bm{\tau}^{\perp}_ \textrm{DNS}$ \\
Solve the RANS equations: $\mathcal{N}\left(\overline{u}^{(i)}\right) = \nabla \cdot \bm{\tau}^{(i)}-\nabla p$ to obtain $\overline{u}^{(i)}$
 }
  \caption{\emph{Implicit} treatment of Reynolds stress that depends on the strain rate among iterations}
  \label{alg:eddy-viscosity-model}
\end{algorithm}
\end{minipage}
\vspace{1em}

\begin{figure}
  \centering
  \hspace{2em}\includegraphics[width=0.4\textwidth]{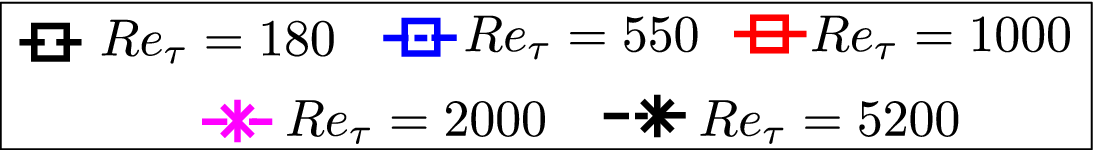}\\
  \includegraphics[width=0.45\textwidth]{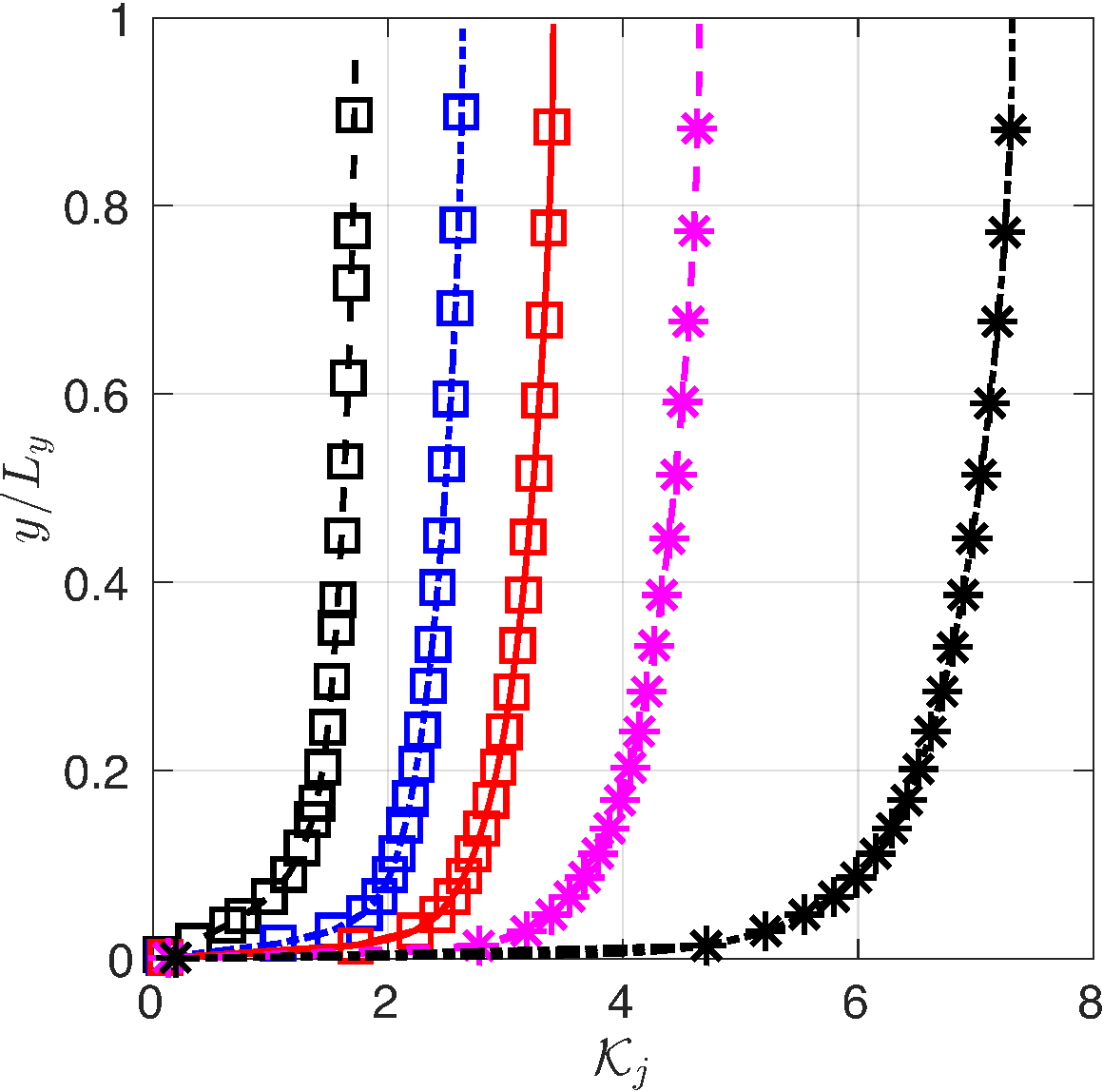}
    \caption{The local condition number at different Reynolds numbers by using \textbf{implicit treatment} of Reynolds stress, i.e., the linear part of Reynolds stress is implicitly treated by introducing an optimal eddy viscosity.}
  \label{fig:cmptKappaTau-nut}
\end{figure}

\begin{figure}
  \centering
   \hspace{1em}\includegraphics[width=0.4\textwidth]{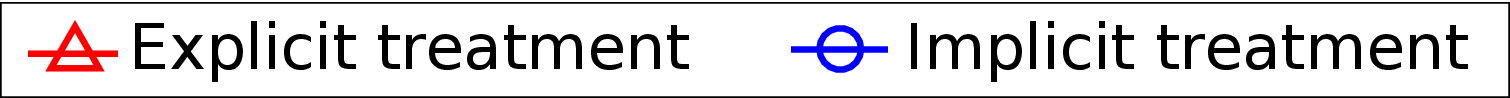} \\
   \includegraphics[width=0.45\textwidth]{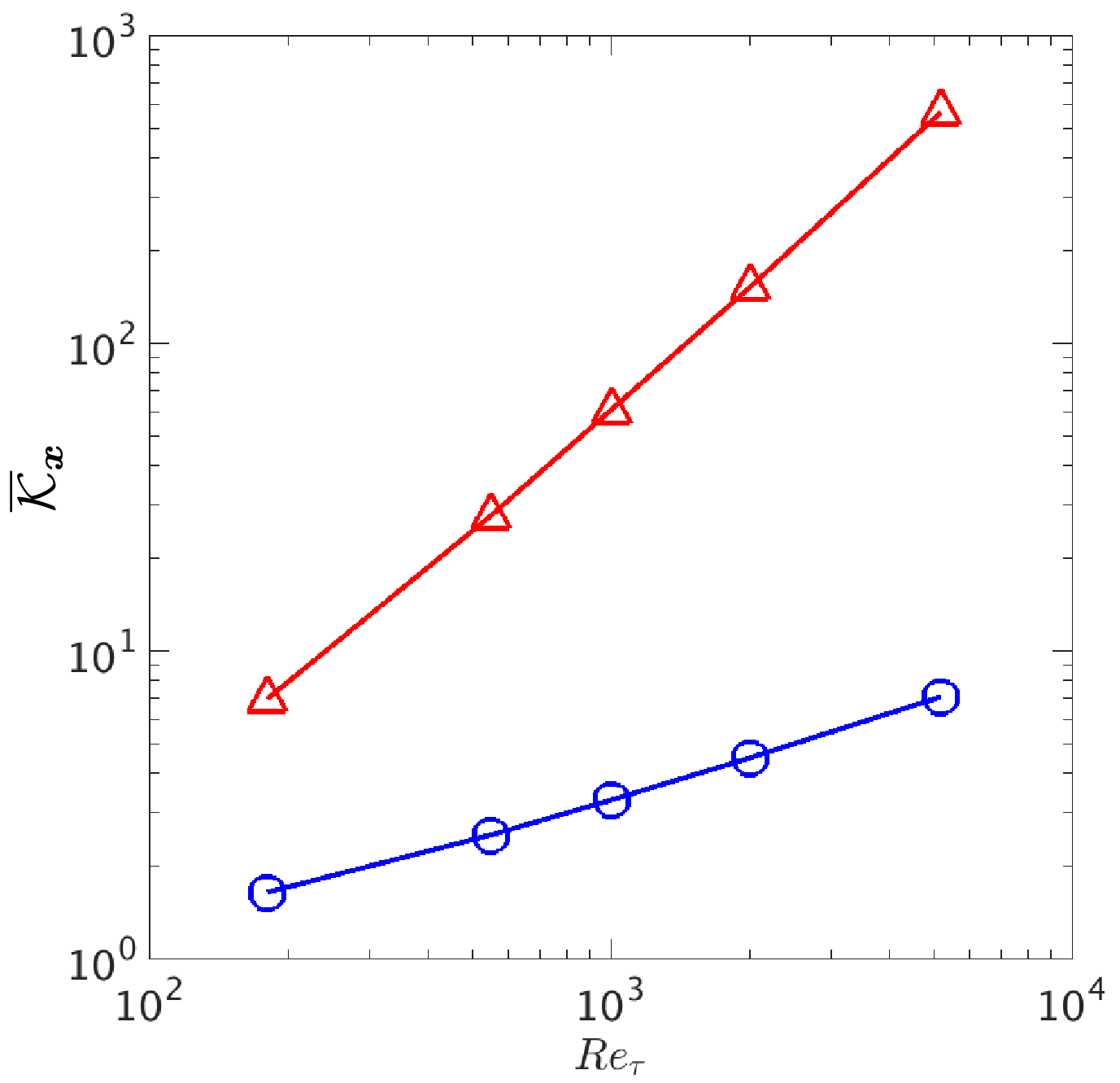} 
    \caption{The volume-averaged local condition number at different Reynolds numbers for explicit and implicit treatments of Reynolds stress.}
  \label{fig:avg-cmptKappaTau-nut}
\end{figure}

Compared to the Reynolds stress models, it is well known that eddy viscosity models can enhance the stability and conditioning of RANS equations with turbulence closures. In the practice of traditional RSM, the modeled Reynolds stress is empirically blended with the Reynolds stress from eddy viscosity models to achieve better convergence and conditioning~\citep{basara03new,maduta17improved}. In this work, we demonstrate that the local condition number $\mathcal{K}_j$ can quantitatively explain the improved conditioning of implicit treatment of Reynolds stress by introducing an eddy viscosity. It can be seen in Fig.~\ref{fig:cmptKappaTau-nut} that the local condition number $\mathcal{K}_j$ is significantly reduced compared with the results of explicit treatment of Reynolds stress as shown in Fig.~\ref{fig:cmptKappaTau}, especially for high Reynolds number flows. Although the local condition number of high Reynolds number is still greater than the one of low Reynolds number, they are at the same order of magnitude for different Reynolds numbers. The volume-averaged local condition number in Fig.~\ref{fig:avg-cmptKappaTau-nut} is also significantly reduced by using implicit treatment of Reynolds stress, demonstrating the merit of using implicit treatment of Reynolds stress in improving the conditioning when solving RANS equations for mean velocity field. The conditioning can be further improved by adjusting $\tau^\perp_{DNS}$ to enhance the implicit treatment part, e.g., increasing $\nu_t^m$ by $\Delta \nu_t$ and adjust nonlinear part of Reynolds stress as $\bm{\tau}^{\perp}_ \textrm{DNS}-\Delta \nu_t \left(\nabla \overline{u}^{(i)}+(\nabla \overline{u}^{(i)})^T\right)$ accordingly. However, it should be noted that such a purely numerical enhancement may introduce excessive errors to iterative solvers when the chosen $\Delta \nu_t$ is too large. The proposed scheme aims to strike a balance between accuracy and conditioning.

\begin{figure}
  \centering
  \subfloat[Relative error]{\includegraphics[height=0.35\textwidth]{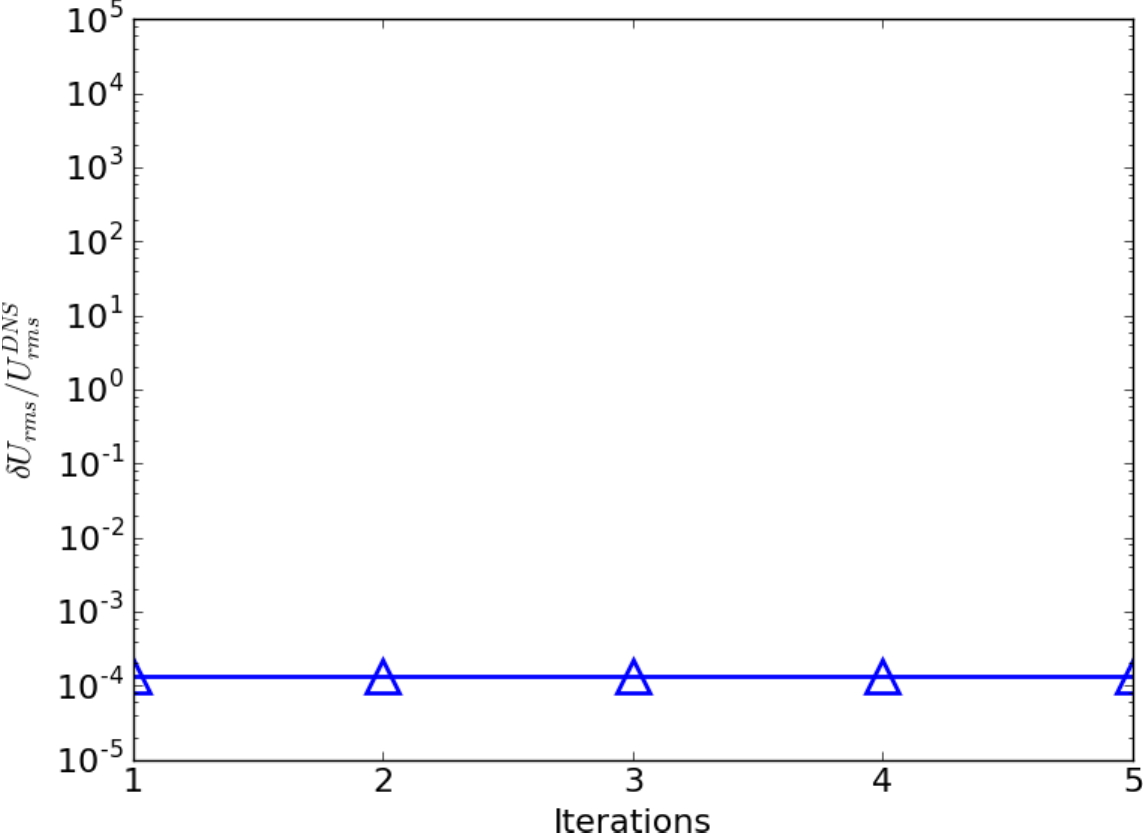}}\hspace{0.5em}
  \subfloat[Condition number]{\includegraphics[height=0.35\textwidth]{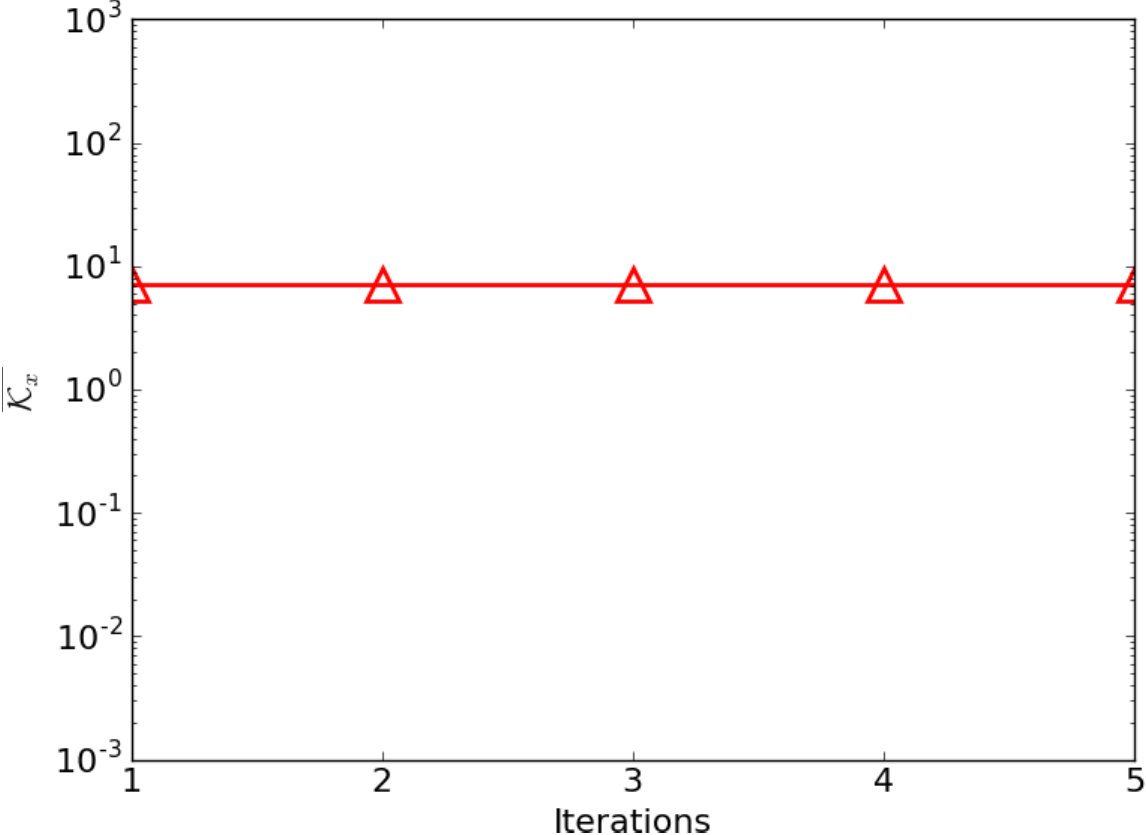}}
    \caption{The error propagation analysis of solving stream-wise velocity iteratively by using \textbf{implicit treatment} of Reynolds stress, including (a) relative error of mean velocity and (b) volume-averaged local condition number.}
  \label{fig:coupled-imp-Ux}
\end{figure}

We further show that the relative error of mean velocity is much smaller by using implicit treatment of Reynolds stress in RANS simulations. It can be seen in Fig.~\ref{fig:coupled-imp-Ux}a that the relative error of the solved mean velocity is much smaller than the one shown in Fig.~\ref{fig:frozen-Ux}a. In addition, the volume-averaged local condition number stays at $O(1)$ as shown in Fig.~\ref{fig:coupled-imp-Ux}b, which explains the better convergence of solving for mean velocity field by using implicit treatment of Reynolds stress.

\begin{figure}
  \centering
  \hspace{2em}\includegraphics[width=0.6\textwidth]{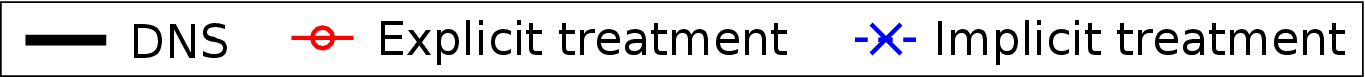}\\
  \subfloat[Mean velocity $U$ ($Re_\tau=180$)]{\includegraphics[width=0.45\textwidth]{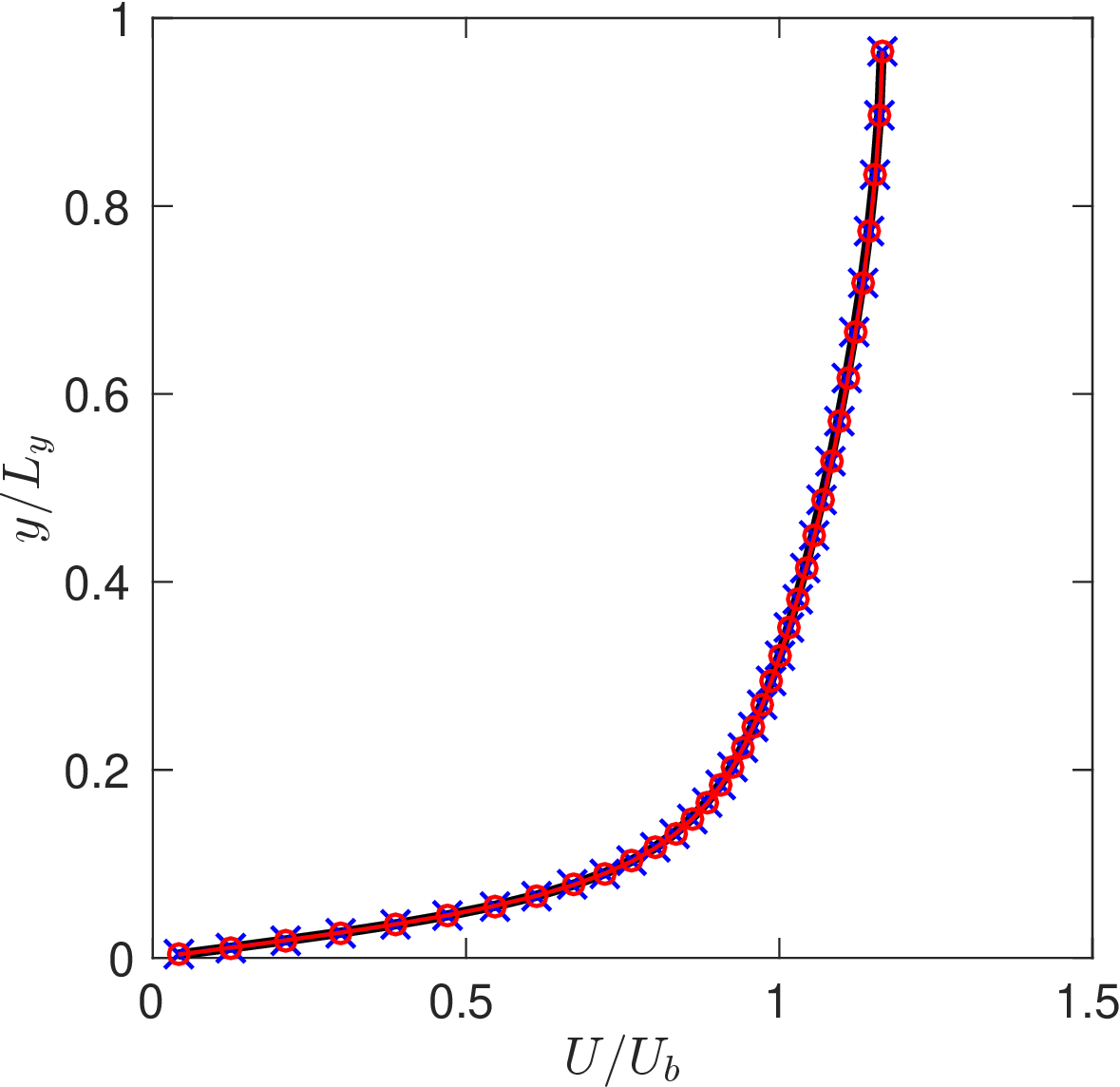}}
  \subfloat[Mean velocity $U$ ($Re_\tau=5200$)]{\includegraphics[width=0.45\textwidth]{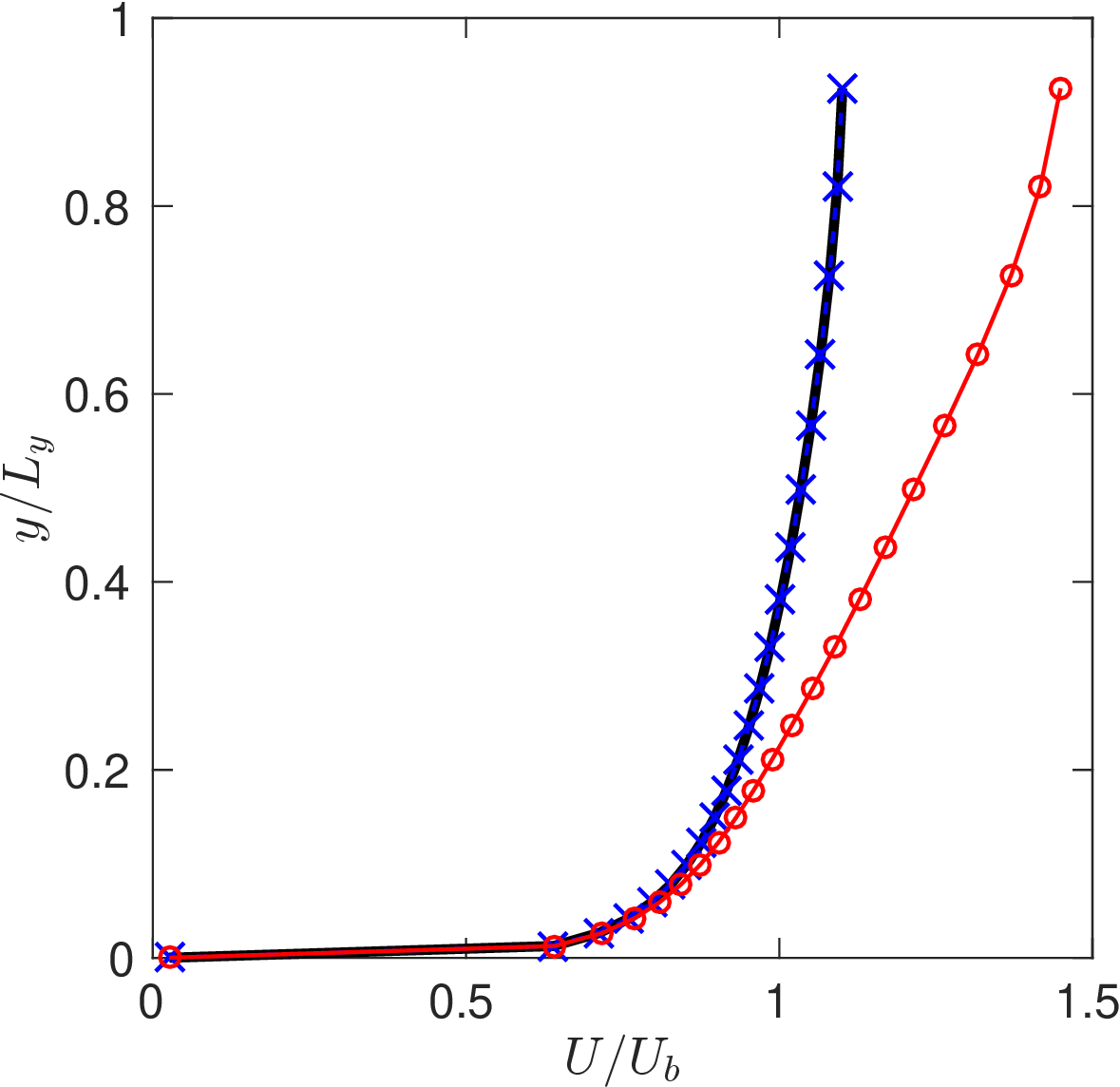}}\\
  \vspace{0.5em}
  \hspace{1.5em}\includegraphics[width=0.47\textwidth]{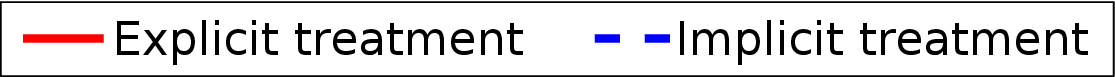}\\
  \subfloat[Percentage error of $U$ ($Re_\tau=180$) ]{\includegraphics[width=0.45\textwidth]{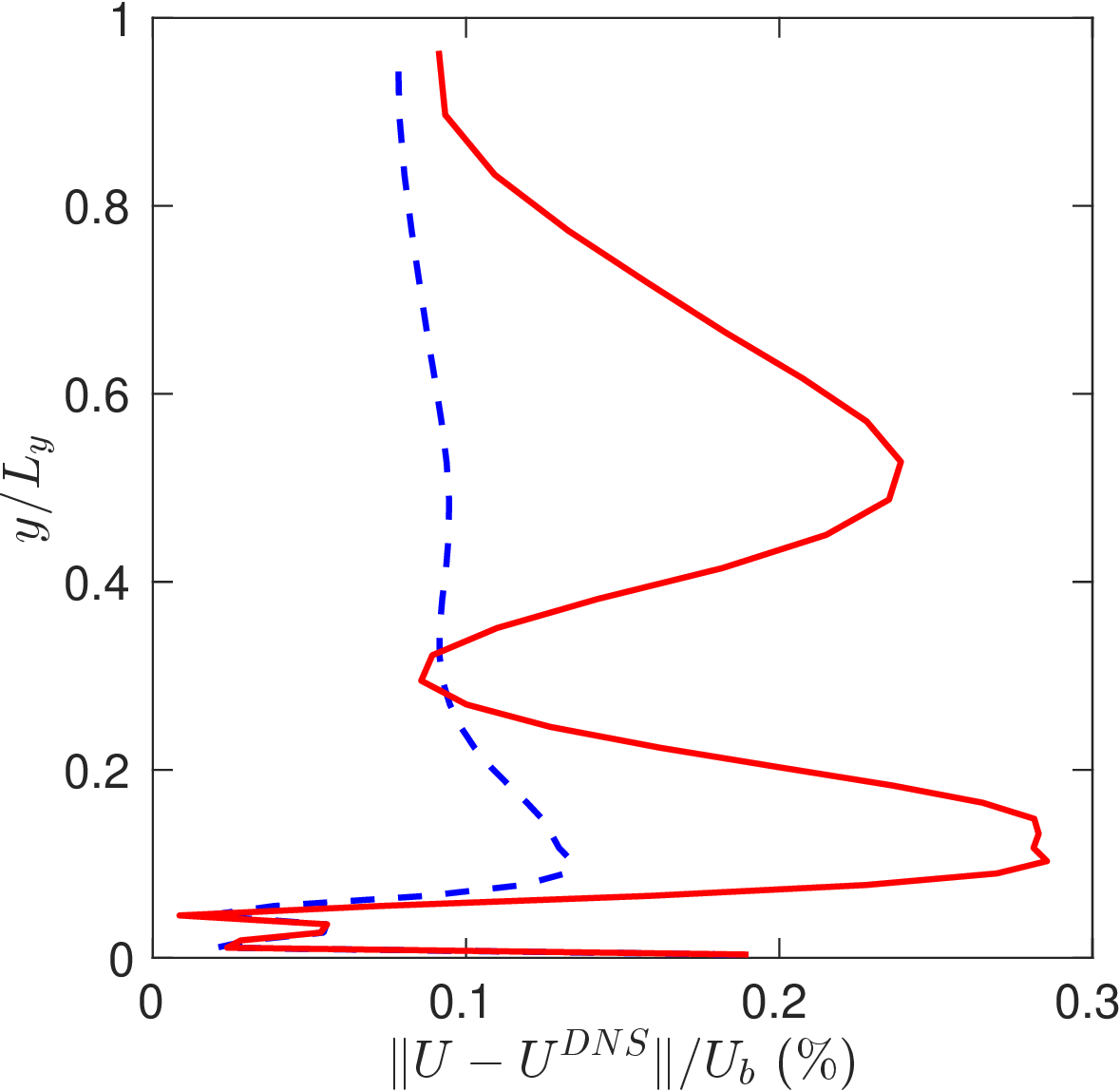}}
  \subfloat[Percentage error of $U$ ($Re_\tau=5200$) ]{\includegraphics[width=0.45\textwidth]{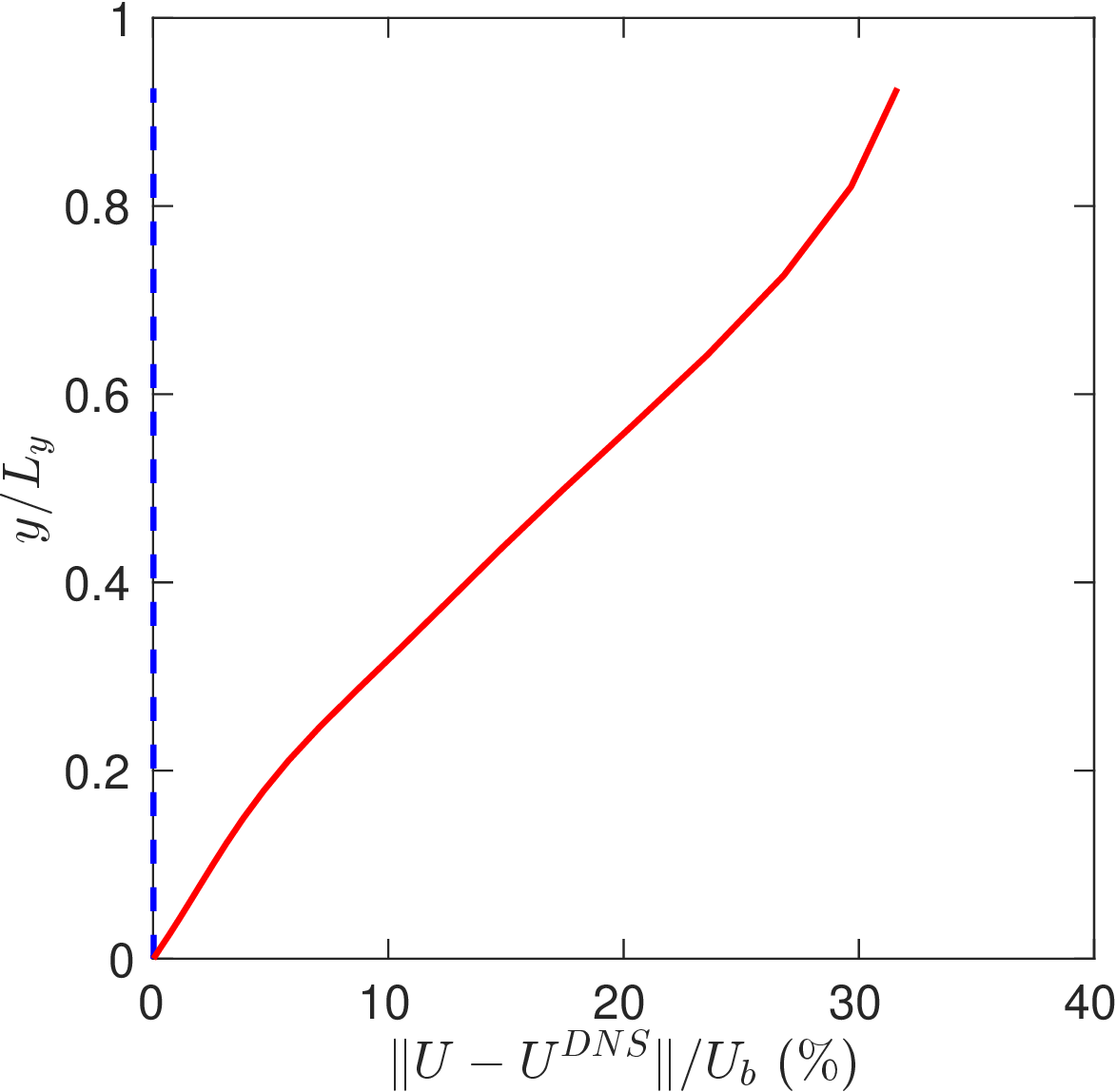}} \\
    \caption{The comparison of solved mean velocity by using explicit and implicit treatments of Reynolds stress, including (a) mean velocity $U$ at $Re_\tau=180$, (b) percentage error of mean velocity $U$ at $Re_\tau=180$, (c) mean velocity $U$ at $Re_\tau=5200$ and (d) percentage error of mean velocity $U$ at $Re_\tau=5200$.}
  \label{fig:U-comp-nut}
\end{figure}

The mean velocity $U$ is solved and presented in Fig.~\ref{fig:U-comp-nut} at Reynolds numbers $Re_\tau=180$ and $Re_\tau=5200$ by using explicit and implicit treatments of Reynolds stress. At Reynolds number $Re_\tau=180$, it can be seen in Fig.~\ref{fig:U-comp-nut}a that the solved mean velocity $U$ by using both kinds of treatments has a good agreement with DNS data. It demonstrates that the error propagation from Reynolds stress to mean velocity is not severe at low Reynolds number, and the percentage error of mean velocity is comparable by using either explicit or implicit treatment of Reynolds stress as shown in Fig.~\ref{fig:U-comp-nut}c. These results have a good agreement with the local condition number presented in Figs.~\ref{fig:cmptKappaTau} and~\ref{fig:cmptKappaTau-nut}, which shows that the local condition number is of the same order for the flow at Reynolds number $Re_\tau=180$ by using both types of treatments. However, the solved mean velocity fields are noticeably different at high Reynolds number ($Re_\tau=5200$) as shown in Fig.~\ref{fig:U-comp-nut}b. Specifically, the solved mean velocity by using explicit treatment of Reynolds stress shows a significant difference from the DNS data, while the solved mean velocity by using implicit treatment of Reynolds stress still agrees well with the DNS data at $Re_\tau=5200$.  The subtle difference between the explicit and implicit treatments are further discussed in Appendix~\ref{sec:app-discuss-paradox}.  The percentage error of solved mean velocity at $Re_\tau=5200$ in Fig.~\ref{fig:U-comp-nut}d also confirms that the error in mean velocity by using explicit treatment of Reynolds stress is orders of magnitude higher than the error of using implicit treatment of Reynolds stress. Such comparison of solved mean velocity fields agrees well with the differences in local condition number $\mathcal{K}_j$, demonstrating that the proposed local condition number can be used to quantitatively assess the error propagation from Reynolds stress to mean velocity when solving RANS equations. At a high Reynolds number ($Re_\tau=5200$), the profile of the error is dominated by the local condition number, which is much larger near the channel center as shown in Fig.~\ref{fig:cmptKappaTau}. At a low Reynolds number ($Re_\tau=180$), the local condition number is of the same order of magnitude in the whole domain, and thus the profile of the error is dominated by the error within the DNS Reynolds stress. Therefore, the noisy pattern of the profile of the error can only be seen in Fig.~\ref{fig:U-comp-nut}c for $Re_\tau=180$ but not in Fig.~\ref{fig:U-comp-nut}d for $Re_\tau=5200$. Inspired by the comparison of conditioning with explicit and implicit treatment of Reynolds stress closures, \cite{wu18data-driven} further proposed an implicit treatment of Reynolds stress for machine-learning-assisted RANS modeling to improve the conditioning when solving for mean velocity field.

\subsection{Model conditioning of RANS equations for more complex flows}
\begin{figure}
\centering
\subfloat[Flow in a square duct]{\includegraphics[height=0.25\textwidth]{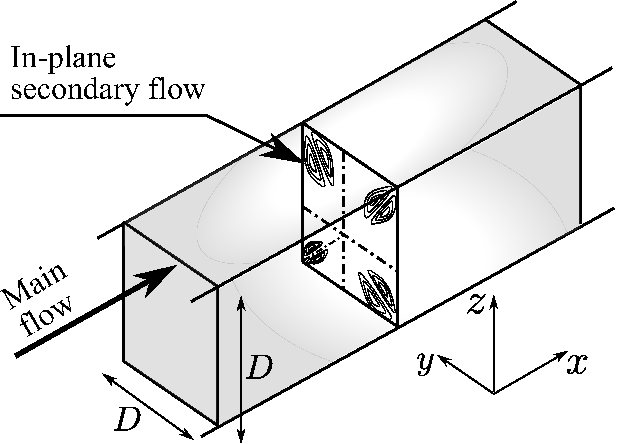}}\hspace{1em}
\subfloat[Flow over periodic hills]{\includegraphics[height=0.25\textwidth]{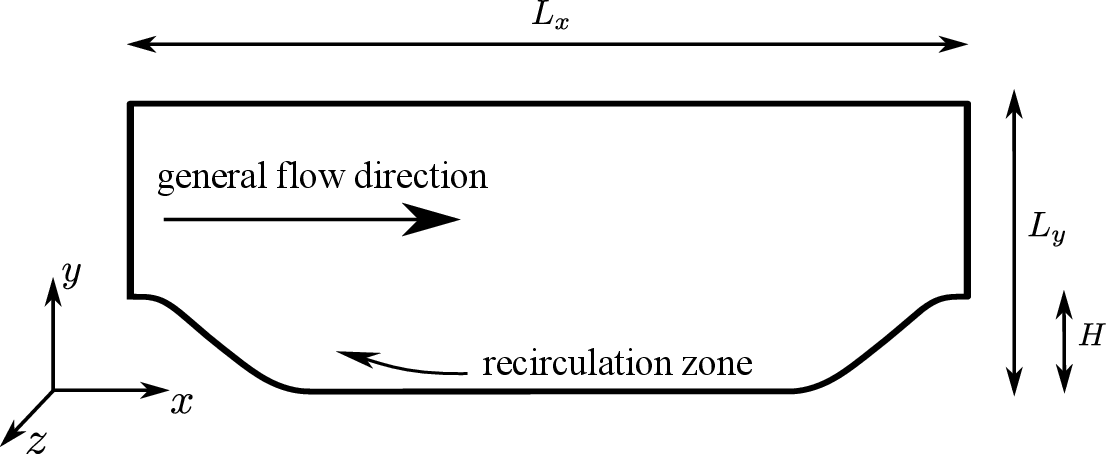}}
\caption{The configuration of (a) the flow in a square duct at $Re_b=3500$ and (b) the flow over periodic hills at $Re_b=5600$.}
\label{fig:flow-config}
\end{figure}

We further study the model conditioning of RANS equations for more complex flows including the flow in a square duct at $Re_b=3500$~\citep{pinelli10reynolds} and the flow over periodic hills at $Re_b=5600$~\citep{breuer09flow}, where $Re_b$ is defined by the bulk velocity $U_b$ at the inlet. The flow configurations are shown in Fig.~\ref{fig:flow-config}. In this work, we show that the RANS equations of more complex flows can also be ill-conditioned. In addition, we demonstrate that the proposed local condition number can be used to assess the model conditioning of RANS equations in these more complex flows.

\subsubsection{Flow in a square duct}
\label{sec:res-duct}
\begin{figure}
\centering
\hspace{1em}\includegraphics[width=0.5\textwidth]{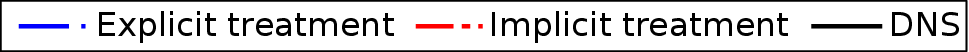}\\
\includegraphics[height=0.35\textwidth]{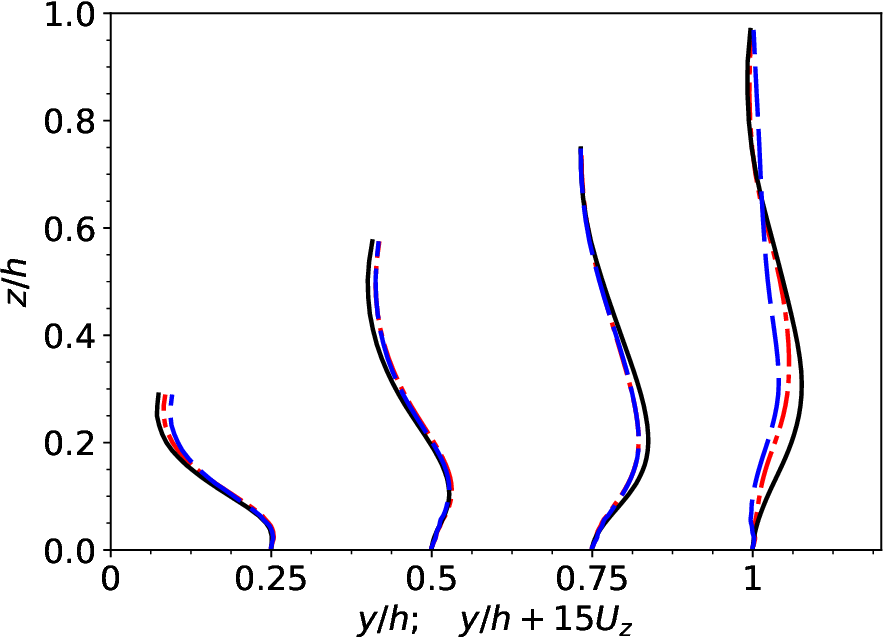}
\caption{The comparison of solved mean velocities by using explicit treatment with fixed Reynolds stress and implicit treatment of Reynolds stress. The computational domain covers a quarter of the cross-section of the physical domain, i.e., $h=D/2$. This is due to the symmetry of the mean flow in both $y$ and $z$ directions as shown in Fig.~\ref{fig:flow-config}a. It should be noted that the Reynolds stress is obtained from DNS database~\citep{pinelli10reynolds}.}
\label{fig:duct-Uz-comp}
\end{figure}

\begin{figure}
\centering
\includegraphics[width=0.3\textwidth]{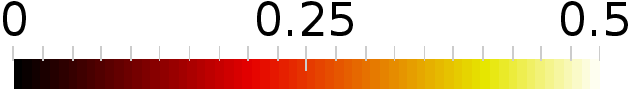}\\
\subfloat[Explicit treatment]{\includegraphics[height=0.35\textwidth]{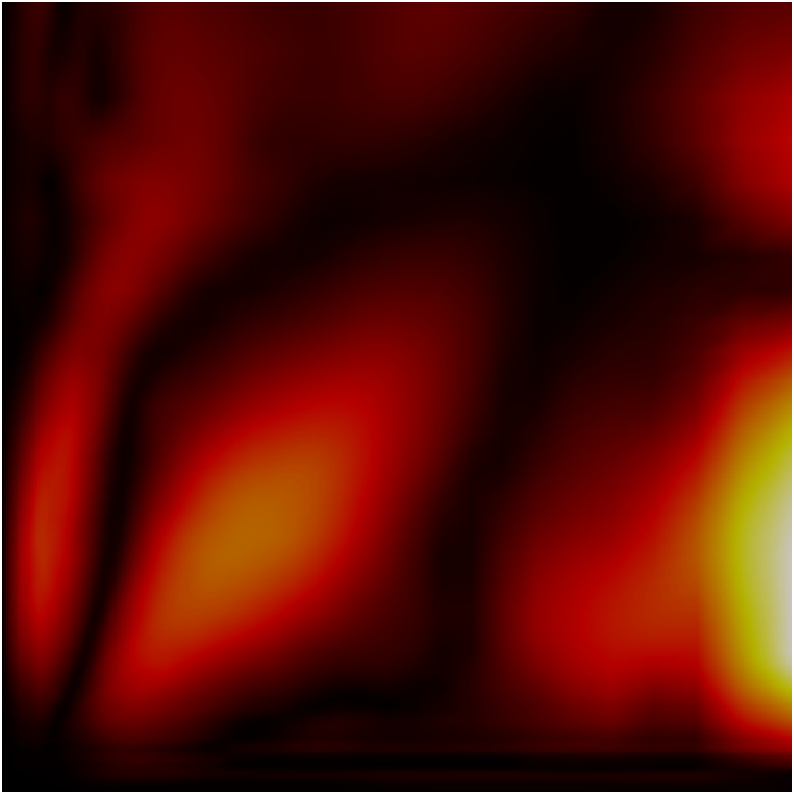}}\hspace{1em}
\subfloat[Implicit treatment]{\includegraphics[height=0.35\textwidth]{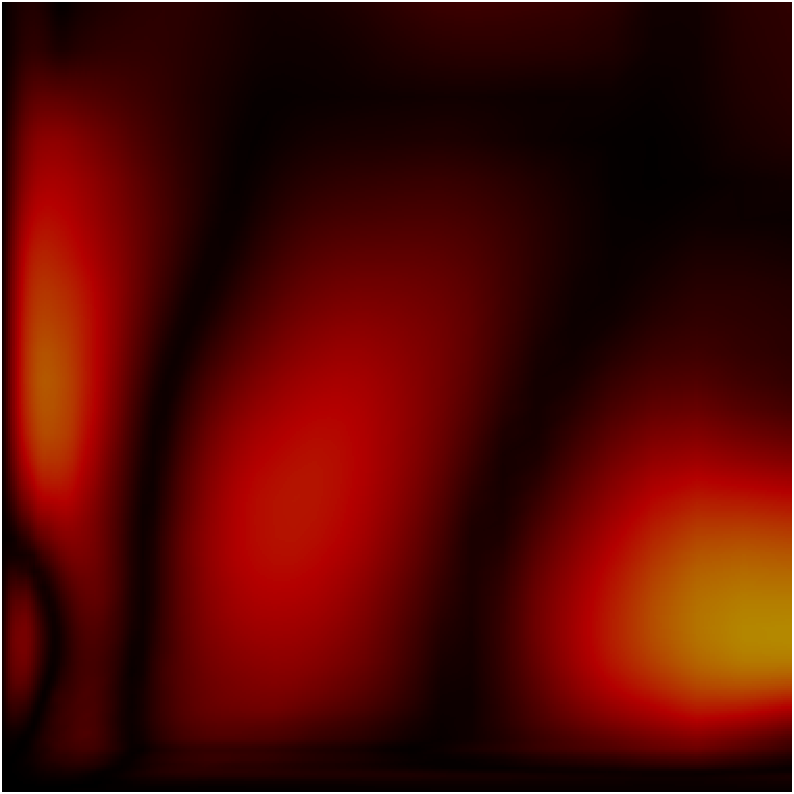}}
\caption{The normalized error of the solved secondary flow velocity $U_z$ of the flow in a square duct by using (a) explicit treatment with fixed Reynolds stress and (b) implicit treatment of Reynolds stress.}
\label{fig:duct-Uz}
\end{figure}

We first study the solved mean secondary velocity $U_z$ by using explicit treatment with fixed Reynolds stress and implicit treatment of Reynolds stress. The comparison of mean velocity profiles demonstrate that the implicit treatment of Reynolds stress leads to solved mean velocity that has better agreement with DNS data in Fig.~\ref{fig:duct-Uz-comp}. We then focus on the analysis of the errors in this work. The error is quantified by the ratio $\|U_z-U_z^\text{DNS}\|/U_{z, \text{rms}}^\text{DNS}$, where $U_{z,\text{rms}}^\text{DNS}$ is an volume averaged velocity of $U_z^\text{DNS}$ as defined in Eq.~\ref{eq:Urms}. It can be seen in Fig.~\ref{fig:duct-Uz}a that some large errors exist in the region of the vertical symmetry plane and around the diagonal within the cross plane. Compared to the errors as shown in Fig.~\ref{fig:duct-Uz}a, noticeable reduction of errors can be observed in Fig.~\ref{fig:duct-Uz}b, where the implicit treatment of the Reynolds stress is used.

\begin{figure}
\centering
\includegraphics[width=0.3\textwidth]{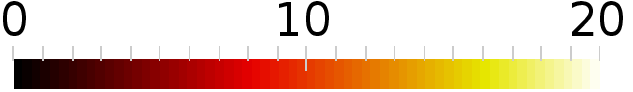}\\
\subfloat[Explicit treatment]{\includegraphics[height=0.35\textwidth]{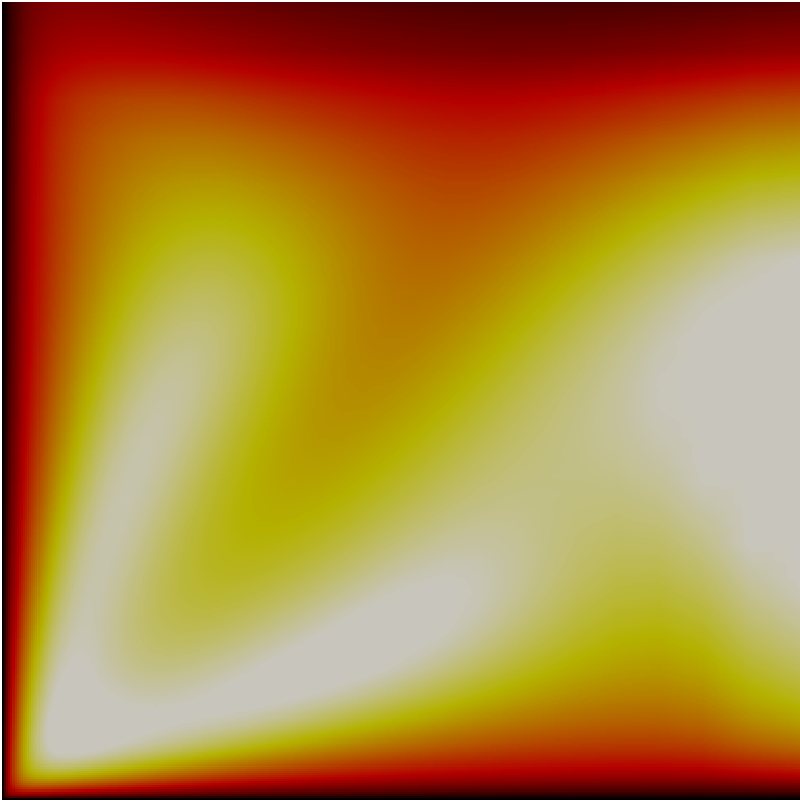}}\hspace{1em}
\subfloat[Implicit treatment]{\includegraphics[height=0.35\textwidth]{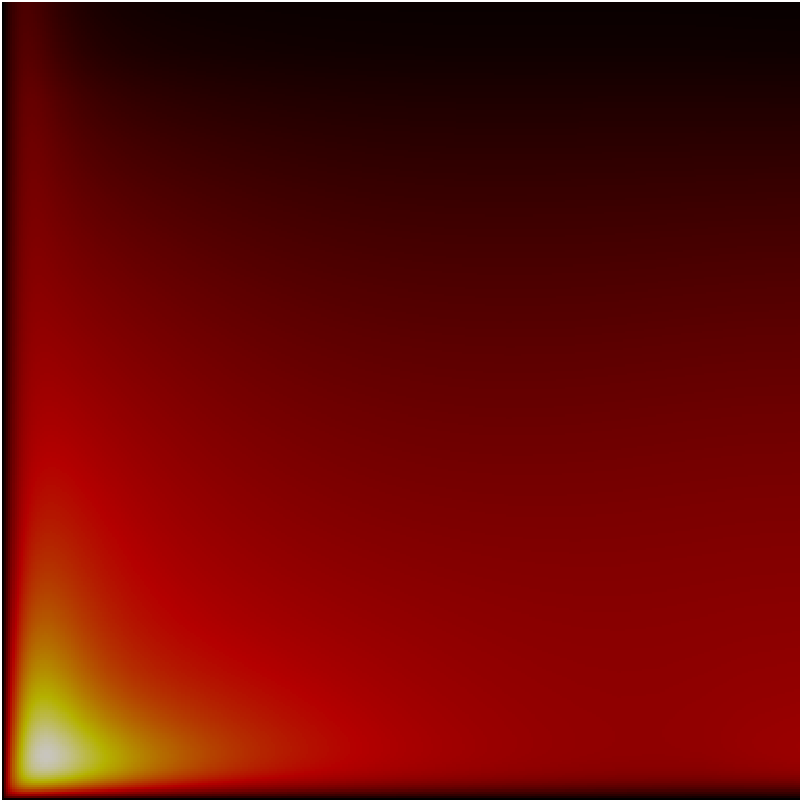}}
\caption{The local condition number of the flow in a square duct by using (a) explicit treatment with fixed Reynolds stress and (b) implicit treatment of Reynolds stress.}
\label{fig:duct-cond}
\end{figure}

The proposed local condition number can be used to analyze the model conditioning for the flow in a square duct as demonstrated in Fig.~\ref{fig:duct-cond}. It can be seen in Fig.~\ref{fig:duct-cond}a that the local condition number is also large in the region of the vertical symmetry plane and around the diagonal within the cross plane, which is consistent with the error of mean velocity as shown in Fig.~\ref{fig:duct-Uz}a. In addition, Fig.~\ref{fig:duct-Uz}b shows that the local condition number is generally smaller across the whole domain by using implicit treatment of Reynolds stress. Such a reduction of local condition number also correlates well with the comparison of mean velocity error in Fig.~\ref{fig:duct-Uz}. It should be noted that the mean velocity error is determined by both the local condition number and the error in Reynolds stress. Therefore, the spatial pattern of mean velocity error in Fig.~\ref{fig:duct-Uz} can not be solely explained by the local condition number in Fig.~\ref{fig:duct-cond}. However, the analysis of local condition number can still provide some information about whether the solved mean velocity is reliable. In practical applications, the error of Reynolds stress is usually unknown, and cautions should be exercised when regions with large local condition number exist. Note that the results shown in Figs.~\ref{fig:duct-Uz} and~\ref{fig:duct-cond} demonstrate that the vertical velocity $U_z$ is not symmetric with itself about the diagonal of the domain. Rather, the diagonal symmetry of this flow is such that the velocity components $U_z$ and $U_y$ are symmetrical to each other about the diagonal. That is, it is expected that $U_y(\mathbf{x}) = U_z(\mathbf{x}')$  and not  $U_z(\mathbf{x}) = U_z(\mathbf{x}')$, where $\mathbf{x}$ and $\mathbf{x}'$ denote two points symmetric about the diagonal.

\subsubsection{Flow over periodic hills}
\label{sec:res-pehill}
\begin{figure}
  \centering
  \hspace{1.5em}\includegraphics[height=0.09\textwidth]{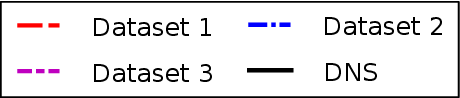}\\
  \subfloat[Reynolds stresses]{\includegraphics[width=0.7\textwidth]{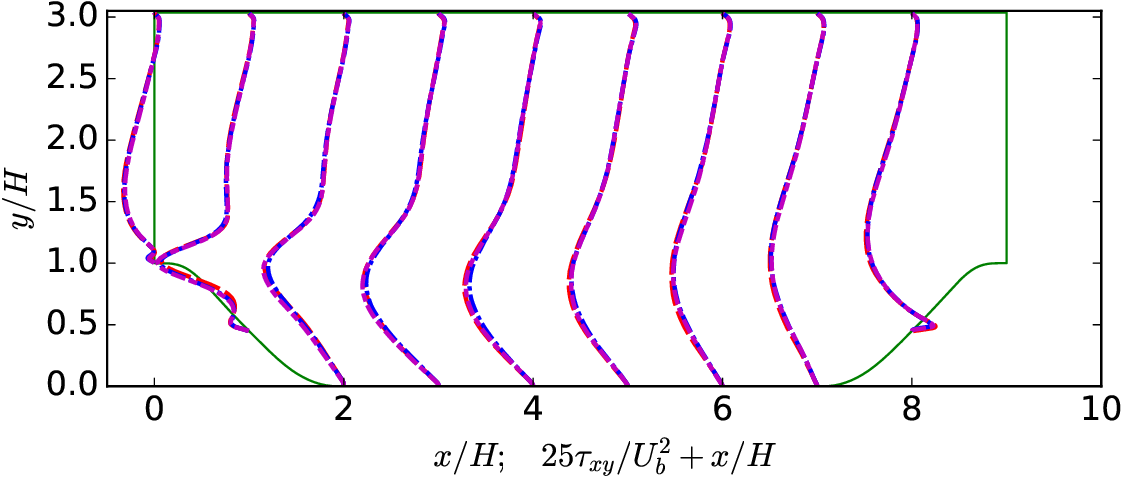}}\hspace{0.5em}
  \subfloat[Mean velocities]{\includegraphics[width=0.7\textwidth]{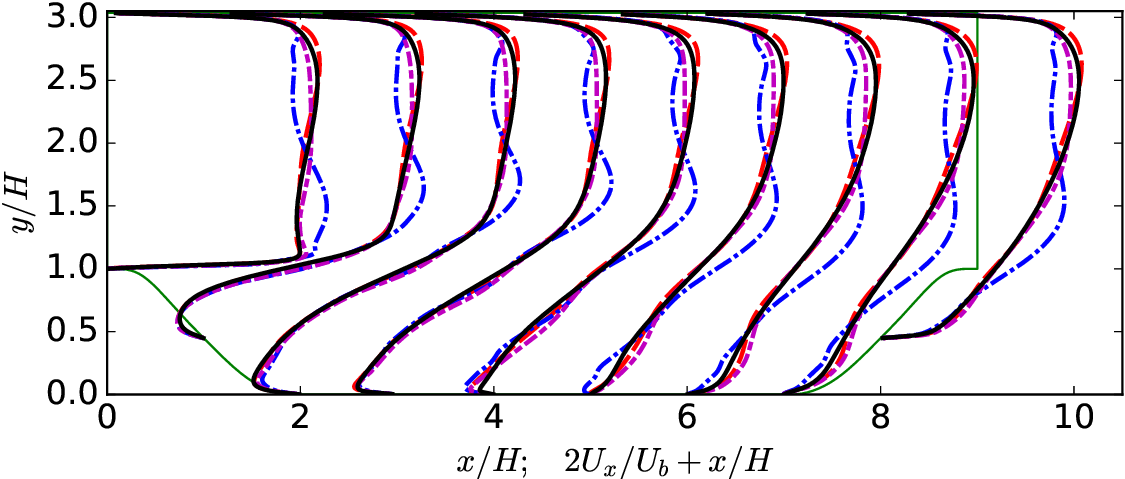}}
    \caption{The comparison of (a) DNS Reynolds stress and (b) solved mean velocity between different DNS datasets. The solved mean velocity is obtained by substituting the corresponding DNS Reynolds stress and solving RANS equations. The solid black lines in panel (b) denote the DNS mean velocity as a benchmark. It should be noted that the DNS mean velocities from different datasets have no noticeable difference, and here we presented the DNS mean velocity from \cite{breuer09flow}.}
  \label{fig:pehill-U-comp}
\end{figure}

Unlike the flow in a square duct, the error of mean velocity can be large across the whole domain for the flow over periodic hills. We first show the comparison of Reynolds stress from different DNS datasets, including the data by \cite{breuer09flow} (Dataset 1) and two datasets obtained by using Incompact3d~\citep{laizet2009high,laizet2011incompact3d} by using two different mesh resolutions (Datasets 2 and 3). The simulation details are summarized in Table~\ref{tab:pehill-case-summary}. It can be seen in Fig.~\ref{fig:pehill-U-comp}a that the Reynolds stresses obtained from these three datasets are very close to each other. According to the comparison of these Reynolds stresses, we might intuitively expect that the solved mean velocity field is similar to each other by substituting the fixed DNS Reynolds stresses into RANS equations. However, it is not the case as shown in Fig.~\ref{fig:pehill-U-comp}b, where we compare the solved mean velocity with the DNS mean velocity field. It can be seen that the solved mean velocity field by using fixed DNS Reynolds stress from dataset 2 show noticeable differences across the whole domain. In contrast, the solved mean velocity fields from datasets 1 and 3 have better agreement with the DNS mean velocity field, but they are still different from each other. It is expected that implicit treatment leads to better agreement of solved mean velocity with DNS data, which has been demonstrated in a related work~\citep{wu18data-driven}.

\begin{table}
  \caption{Summary of the datasets of the flow over periodic hills at Reynolds number $Re=5600$, including the numerical methods, treatment of solid boundaries, accuracy of numerical discretization, mesh sizes and type.
  $Nx$, $N_y$, and $N_z$ indicate number of grids in streawise, wall-normal, and spanwise directions, respectively.
  }
  \begin{center}
  \begin{tabular}[c]{M{1cm}M{2.75cm}M{3cm}M{6cm}}
    \hline
    Dataset & Mesh ($N_x\times N_y\times N_z$)
    & Solver
    & Methods \\
    \hline
	1
	& $281\times234\times200$
	& LESOCC\newline
    \citep{breuer09flow}
	& LES, finite volume method, body-fitting grid, second order discretization 
	 \\
	\hline 
         2  
         & $512\times257\times128$ 
         & \multirow{2}{2.75cm}{\centering 
         Incompact3d
         \citep{laizet2011incompact3d}}
         & \multirow{2}{=}[1mm]{\centering
         DNS,
         Cartesian grid with 
         immersed boundary method,
         pseudo-spectral method, sixth order discretization}
        \\
         \cline{1-2}
         \vspace{0.3em}3
         & \vspace{0.3em} $768\times385\times128$
         &
         & 
          \\
    \hline
  \end{tabular}
 \end{center}
 \label{tab:pehill-case-summary}
\end{table}

\begin{figure}
  \centering
  \includegraphics[width=0.3\textwidth]{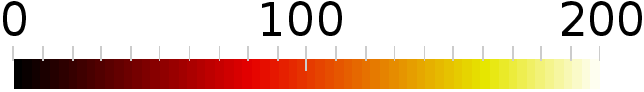}\\
  \subfloat[Explicit treatment]{\includegraphics[width=0.5\textwidth]{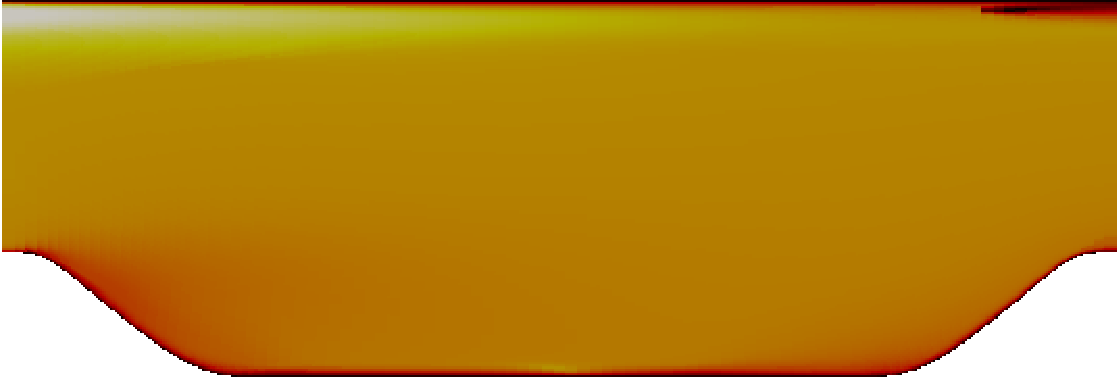}}\hspace{0.5em}
  \subfloat[Implicit treatment]{\includegraphics[width=0.5\textwidth]{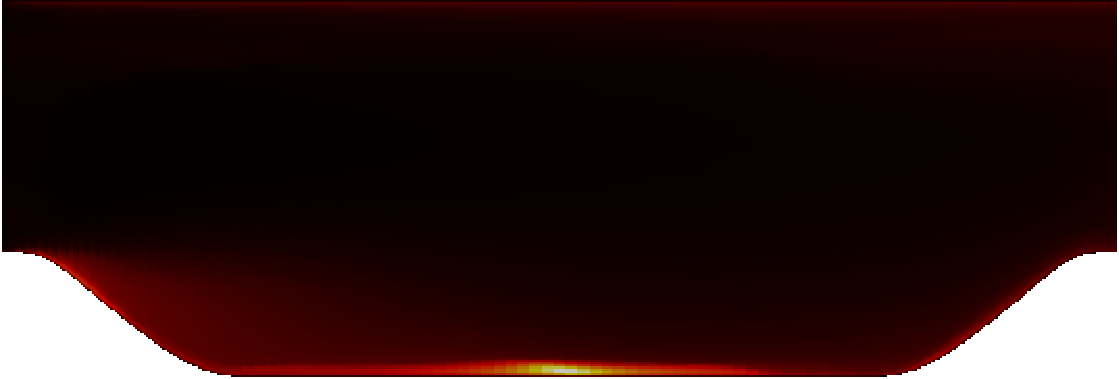}}
    \caption{The local condition number $\mathcal{K}_j$ of the flow over periodic hills at $Re$=5600 by using (a) explicit treatment with fixed Reynolds stress and (b) implicit treatment of Reynolds stress.}
  \label{fig:pehill-cond}
\end{figure}

The comparison in Fig.~\ref{fig:pehill-U-comp} can be explained by studying the model conditioning of RANS equations with specified Reynolds stress. Specifically, it can be seen in Fig.~\ref{fig:pehill-cond}(a) that the local condition number by using fixed Reynolds stress is of the order $O(10^2)$ in most areas, indicating that the RANS equations are ill-conditioned in these regions. On the other hand, the local condition number is smaller in the near-wall and the recirculation regions. The large local condition number can explain the comparison in Fig.~\ref{fig:pehill-U-comp}, i.e., the similar Reynolds stress fields lead to dramatically different mean velocity fields by solving the RANS equations. The physical justification of such a pattern of local condition number field is that the mean velocity error is strongly correlated along the streamlines, and the periofic boundary condition of the inlet and outlet exacerbates the model conditioning for this flow. We also study the local condition number by using implicit treatment of Reynolds stress. It can be seen in Fig.~\ref{fig:pehill-cond}(b) that the local condition number is much smaller than those where fixed Reynolds stress is used in Fig.~\ref{fig:pehill-cond}(a). Therefore, it can be expected that the solved mean velocity by using implicit treatment of DNS Reynolds stress have a better agreement with the DNS mean velocity, which has been confirmed by~\cite{wu18data-driven}.

In this work, the local condition number assess the relative error of solved mean velocity at a given point with regard to the errors in the Reynolds stress field. Therefore, the global effect of error can be captured by the local condition number. Specifically, such an effect depends on the mean flow pattern, which is embodied in the differential operator of linearized RANS equations and influences the local condition number via the Green's function $G({\mathbf{x}}, \bm{\xi})$ as defined in Eq.~(\ref{eq:cont-cond}). For instance, the errors in the solved mean velocity of the periodic hill flow is generally large across the upper channel region as shown in Fig.~\ref{fig:pehill-U-comp}. This is largely due to the fact that the cyclic boundary conditions of the inlet and outlet introduce a strong correlation among errors in the upper channel region. It can be seen in Fig.~\ref{fig:pehill-cond} that the local condition number $\mathcal{K}_j$ is also large across the upper channel region, demonstrating that the local condition number takes into account the mean flow pattern and thus truthfully reflects the potentially large error in the solved mean velocities therein.

\section{Discussions}
\label{sec:discussion}
While this work primarily focuses on the conditioning of a particular class of data-driven Reynolds stress models,
the conditioning issue is an equally important challenge for traditional Reynolds stress models based on Reynolds stress transport equations. Although solving a monolithic system of Reynolds stress transport equations and RANS equations is the most effective way to improve conditioning, it is uncommon due to increased computational costs. 
Moreover, monolithic coupling is by no means a panacea that guarantees well-conditioning and stability. The conditioning and stability ultimately depend on the characteristics of the turbulence model itself. For example, the popularity of S--A model in external aerodynamics is largely attributed to its excellent robustness in terms of both model conditioning and numerical stability, which many other models do not have.
As of now, most commercial and open-source general-purpose CFD packages\citep[e.g.,][]{weller98tensorial} still solve the turbulence transport equations and the RANS mean flow equations in a segregated manner, even in solvers where velocity and pressure are solved concurrently. In such segregated solvers, the modeled Reynolds stress is often updated with the mean velocity field at every iteration step, in the hope that the mean velocity field and the Reynolds stress can consistently adjust to each other during the iterations. However, if the RANS equations are ill-conditioned and the Reynolds stresses are treated explicitly as source terms, the error can be amplified within each iteration. Consequently, a small error in the modeled Reynolds stress can lead to large errors in the solved mean velocity field, which is carried over to the Reynolds stresses in the next iteration step and further amplified. Such an error amplification destabilizes the solution procedure and can potentially lead to divergence.

For the reasons outlined above, RANS simulations with Reynolds-stress-based turbulence models need to be stabilized to increase the robustness of the solvers. Examples of stabilization include (1) using the velocity solved with eddy viscosity models as an initial condition for iterations in the RSM-based solver and (2) partial implicit treatment of the Reynolds stress, among others. In the latter category, researchers introduced a hybrid scheme of computing Reynolds stress by blending the RSM modeled Reynolds stress with that computed from eddy viscosity models, with the later stabilizing the solution~\citep{basara03new,maduta17improved}. However, the choice of the blending factor is largely ad hoc due to the lack of a quantitative method to evaluate the model conditioning. A large blending factor improves the conditioning and stabilizes the solution of the RANS equations, but it impairs the accuracy of the solved mean velocity, since the linear eddy viscosity assumption would be increasingly dominant. The metric proposed in this work can assess the model conditioning with any given blending factor, and thus it is possible to choose a minimum blending factor that maintains good conditioning.

A monolithic coupling for data-driven turbulence models is more challenging than for traditional PDE-based models, if possible at all. For example, for a neural-network-based data-driven turbulence model \citep[e.g.,][]{ling16reynolds,zhang2018machine}, a monolithic coupling is possible, because neural networks models are differentiable. However, for non-differentiable models, e.g., those based on random forests or other tree-based models~\citep[e.g.,][]{wang17physics-informed}, a monolithic coupling is not straightforwardly viable.

\section{Conclusion}
\label{sec:conclusion}
Recently, several researchers employed DNS Reynolds stress data as the closure term and solved the RANS equations for mean velocities on turbulent channel flows. They reported unexpected results that the obtained mean velocities deviated significantly (up to 35\%) from the DNS data at high Reynolds numbers. In this work, we aim to identify a metric to quantitatively assess the conditioning of RANS equations with data-driven Reynolds stress closures, i.e., how a small error in Reynolds stress can lead to large errors in the mean velocity by solving RANS equations. The turbulent channel flow is studied to evaluate the candidate metrics. Our analysis shows that the global, matrix-based condition number is not able to distinguish the different sensitivity of solved mean velocities at different Reynolds numbers. A local condition number function is then derived as a more precise indicator of model conditioning. We demonstrate that such a local condition number explains the error propagation from the modeled Reynolds stress to the solved mean velocity in RANS simulations for turbulent channel flows at different Reynolds numbers. Two more complex flows are also studied to further demonstrate the capability of the proposed local condition number in evaluating the conditioning of RANS equations with data-driven Reynolds stress closures. The proposed condition number provides a quantitative metric to assess the model conditioning of RANS equations, facilitating the development of conditioning-oriented schemes in data-driven turbulence modeling.

\section*{Acknowledgment}
The authors would like to thank Dr. Sylvain Laizet of Imperial College London
and Prof. Michael Breuer of Helmut-Schmidt-Universit\"at Hamburg
for providing us with the high-fidelity simulation data of the flow over periodic hills. The authors would also like to thank Dr.~Gary N.~Coleman of NASA Langley and Dr.~Florian Menter of ANSYS for the helpful discussions during this research.  We gratefully acknowledge Dr.~Scott Murman of NASA Ames and Prof.~Svetlana Poroseva of The University of New Mexico for their constructive comments on the first draft of the manuscript. The authors also would like to thank the reviewers for their constructive and valuable comments, which helped improving the quality and clarity of this manuscript.

\appendix 
\section{Derivations of Condition Numbers}
\subsection{Derivation of global, matrix-based condition number}
\label{app:global-cn}

\noindent
The global matrix-based condition number $\mathcal{K}_{\tau}$ is defined as follow:
\begin{linenomath}
\begin{equation}
\label{eq:app-global-define}
 \frac{\|\delta \mathbf{U}\|}{\|\mathbf{U}\|}  \le 
\mathcal{K}_{\tau} \frac{\|\nabla \cdot \delta \bm{\tau} \|}{\|\nabla \cdot \bm{\tau} \|}
\end{equation}
\end{linenomath}
where $\mathcal{K}_{\tau}$ measures the sensitivity of the solved mean velocity field due to the perturbation of Reynolds stress field, and $|\cdot|$ indicates Euclidean norm of a vector (of all values in the discretized velocity field).  To derive the formulation of $\mathcal{K}_{\tau}$, the perturbation $\delta \bm{b}$ in Eq.~(\ref{eq:global-cn}) is further written as:
\begin{linenomath}
\begin{equation}
\delta \bm{b}  =  \nabla \cdot \delta \boldsymbol{\tau} -
\delta (\nabla p)
\end{equation}
\end{linenomath}
For the purpose of the sensitivity study here, it is assumed that a constant pressure gradient is imposed to drive the flow, i.e., $\delta (\nabla p) = 0$, and thus we have:
\begin{linenomath}
\begin{equation}
  \label{eq:app-delta-b}
  \delta \bm{b} =  \nabla \cdot
  \delta \boldsymbol{\tau}.  
\end{equation}
\end{linenomath}
Hence, 
\begin{linenomath}
\begin{equation}
  \label{eq:app-cond-norm-derive}
 \frac{\|\delta \mathbf{U}\|}{\|\mathbf{U}\|}  \le 
\mathcal{K}_\mathsf{A} \frac{\|\delta \bm{b} \|}{\| \bm{b} \|}   
 = \mathcal{K}_\mathsf{A}  \frac{\| \nabla \cdot  \delta \boldsymbol{\tau} \|}{\| \bm{b} \|}  
= \underbrace{\mathcal{K}_\mathsf{A}  \frac{\|\nabla \cdot \boldsymbol{\tau}\|}{\|\bm{b}\|}}_{\textstyle \mathcal{K}_{\tau}}
\frac{\|\nabla \cdot  \delta \boldsymbol{\tau} \|}{\| \nabla \cdot  \boldsymbol{\tau} \|} .
\end{equation}
\end{linenomath}
Comparing the forms of Eq.~(\ref{eq:app-cond-norm-derive}) with the definition of $\mathcal{K}_\tau$ in Eq.~(\ref{eq:app-global-define}),
the matrix-norm-based condition number for Reynolds-stress-based turbulence models is thus:
\begin{linenomath}
\begin{equation}
  \label{eq:app-cond-norm}
\mathcal{K}_\tau = \mathcal{K}_\mathsf{A} \frac{\|\nabla \cdot
  \boldsymbol{\tau}\|}{\|\bm{b}\|}
\end{equation}
\end{linenomath}

\subsection{Derivation of local condition number function}
\label{sec:app-local-cn}
\noindent
The continuous local condition number $\mathcal{K}(\mathbf{x})$ is defined as follow:
\begin{linenomath}
\begin{equation}
\label{eq:app-define-Kx}
 \frac{|\delta \bm{u}(\bm{x})|}{U_{\infty}}  \le 
\mathcal{K}(\bm{x}) \frac{\Cnorm{\nabla \cdot \delta \bm{\tau}}}{\Cnorm{\nabla \cdot \bm{\tau}}}
\end{equation}
\end{linenomath}
where $\mathcal{K}(\bm{x})$ measures the sensitivity of the solved mean velocity at any given location $\bm{x}$ due to the perturbation of the Reynolds stress field, and $U_\infty$ is a constant representative velocity magnitude for normalization. The function norm $\Cnorm{\cdot}$ of function $f(\ms{\xi})$ on domain $\Omega$ is defined as in Eq.~(\ref{eq:function-norm-def}).

To derive the formulation of this local condition number, we first consider the solution $\bm{u}$ at a particular location $\mb{x}'$:
\begin{linenomath}
\begin{equation}
\label{eq:app-u0}
\bm{u}(\mb{x}') = \int_\Omega G(\mb{x}'; \ms{\xi}) \, \mb{b}(\ms{\xi}) \, d\ms{\xi}
\end{equation}
\end{linenomath}
where $G$ represents the Green's function of the linear differential operator $\mathcal{L}$ in the linearized RANS equations as defined in Eq.~(\ref{eq:L-definition}). Denoting $G_{\mb{x}'} = G(\mb{x}'; \ms{\xi})$,
the perturbation of the solution is thus:
\begin{linenomath}
\begin{equation}
\delta \bm{u}(\mb{x}') = \int_\Omega G(\mb{x}'; \ms{\xi}) \, \delta\mb{b}(\ms{\xi}) \, d\ms{\xi} = \Cinner{G_{\mb{x}'},\delta\mb{b}} 
\end{equation}
\end{linenomath}
where $\Cinner{\cdot}$ is the inner product of functions defined on domain $\Omega$.

Using the Schwartz inequality~\citep{steele04cauchy,debnath05hilbert} leads to:
\begin{linenomath}
\begin{align}
|\delta \bm{u}(\mb{x}')| & \le
\Cnorm{G_{\mb{x}'}} \,
\Cnorm{\delta \mb{b}}  \\
& = \Cnorm{G_{\mb{x}'}} \,
\Cnorm{\nabla \cdot \delta \ms{\tau}} 
\end{align}
\end{linenomath}
As in~\ref{app:global-cn}, the pressure gradient is assumed constant and thus $\delta \bm{b} =  \nabla \cdot \delta \boldsymbol{\tau}$. Finally,  the sensitivity of mean velocity $\bm{u}$ with respect to the Reynolds stress $\bm{\tau}$ perturbations is derived as follows:
\begin{linenomath}
\begin{align}
\frac{|\delta \bm{u}(\mb{x}')|}{U_\infty} 
& \le \frac{\Cnorm{G_{\mb{x}'}} \,
\Cnorm{\nabla \cdot \delta \ms{\tau}}}{U_\infty} \\
& = 
\frac{\Cnorm{G_{\mb{x}'}} \,
\Cnorm{\nabla \cdot \ms{\tau}}}{U_\infty} \,
\frac{\Cnorm{\nabla \cdot \delta \ms{\tau}}}
{\Cnorm{\nabla \cdot \ms{\tau}}}
\label{eq:app-cond-derive}
\end{align}
\end{linenomath}
Therefore, by comparing Eqs.~(\ref{eq:app-cond-derive}) and~(\ref{eq:app-define-Kx}), we define a \textbf{local condition number function} $\mathcal{K}$ of spatial location $\mb{x}$ as:
\begin{linenomath}
\begin{equation}
\label{eq:app-cont-cond}
\mathcal{K}(\mb{x}) = 
\frac{\Cnorm{G_{\mb{x}}} \,
\Cnorm{\nabla \cdot \ms{\tau}}}{U_\infty} 
= \frac{
\Cnorm{G(\mb{x}, \bm{\xi})} \,
\Cnorm{\nabla \cdot \ms{\tau}}
}{U_\infty}
\end{equation}
\end{linenomath}
Without causing ambiguity, we have dropped the subscript of $\bm{x}'$ in the equation above and in the text for simplicity of notation.

\subsection{Local condition number for implicit treatment of Reynolds stress}
\label{sec:app-eddy-cn}
\begin{figure}
\centering
\hspace{1.5em}\includegraphics[width=0.9\textwidth]{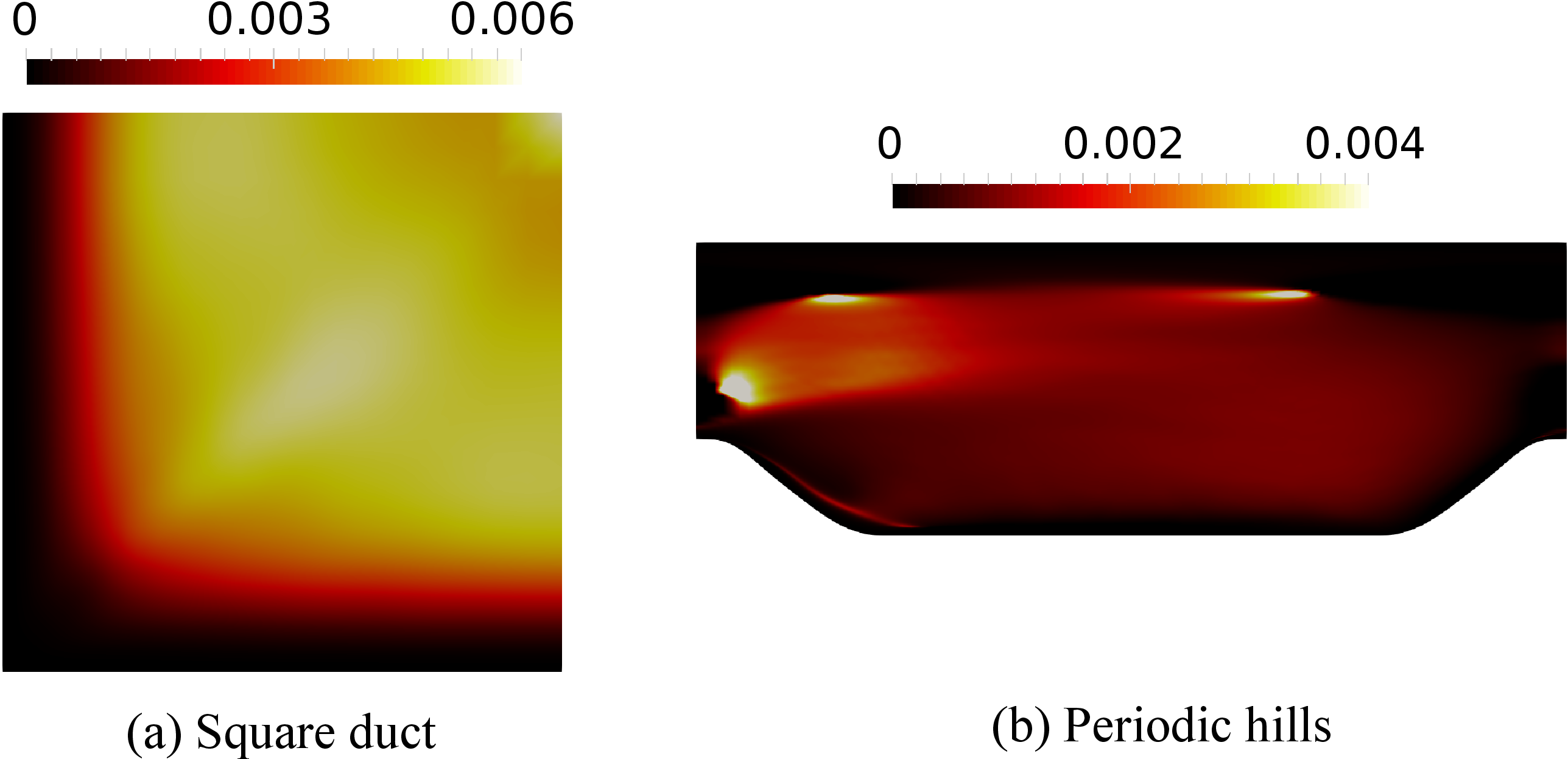}
\caption{The optimal eddy viscosity fields for (a) the flow in a square duct at $Re_b=3500$ and (b) the flow over periodic hills at $Re_b=5600$.}
\label{fig:nut}
\end{figure}

In the practice of RANS modeling, eddy viscosity models are widely used, and the modeled eddy viscosity influences the differential operator $\mathcal{L}$ associated with RANS equations. Therefore, we extend the derivation of Eq.~(\ref{eq:cont-cond}) to make it compatible with the implicit treatment of Reynolds stress. According to the general form of implicit treatment of Reynolds stress~\citep{pope75more} in Eq.~(\ref{eq:nut-model}), the linearized RANS equations in Eq.~\ref{eq:ns-concise-L} can be rearranged as follow:
\begin{linenomath}
\begin{equation}
  \label{eq:app-ns-concise-nut}
  \widetilde{\mathcal{L}}(\bm{u}) = \nabla \cdot \boldsymbol{\tau}^\perp-\nabla p
\end{equation}
\end{linenomath}
where $\widetilde{\mathcal{L}}=\mathcal{L}-\nu_t^m \nabla^2$ is the modified linear differential operator by using implicit treatment of Reynolds stress. Examples of optimal eddy viscosity $\nu_t^m$ for the flow over periodic hills and the flow in a square duct are shown in Fig.~\ref{fig:nut}.Here we only study the perturbation on the nonlinear term $\bm{\tau}^\perp$ of Reynolds stress $\bm{\tau}$, i.e.,
\begin{linenomath}
\begin{equation}
\delta \ms{\tau} = \delta \ms{\tau}^{\perp}
\end{equation}
\end{linenomath}

Finally, we have the local condition number $\mathcal{K}(\mb{x}')$ in Eq.~(\ref{eq:cont-cond}) re-derived as follows for the implicit treatment of Reynolds stress:
\begin{linenomath}
\begin{align}
\frac{|\delta \bm{u}(\mb{x}')|}{U_\infty} 
& \le \frac{\Cnorm{\widetilde{G}_{\mb{x}'}} \,
\Cnorm{\nabla \cdot \delta \ms{\tau}^\perp}}{U_\infty} \\
& = 
\underbrace{\frac{\Cnorm{\widetilde{G}_{\mb{x}'}} \,
\Cnorm{\nabla \cdot \ms{\tau}}}{U_\infty}}_{\textstyle \mathcal{K}_j} \;\;
\frac{\Cnorm{\nabla \cdot \delta \ms{\tau}}}
{\Cnorm{\nabla \cdot \ms{\tau}}}
\end{align}
\end{linenomath}
where the kernel function $\Cnorm{\widetilde{G}_{\mb{x}'}}$ represents the Green's function that corresponds to the linear differential operator $\widetilde{\mathcal{L}}$ defined in Eq.~(\ref{eq:L-implicit}), taking into account the implicit modeling of linear part of Reynolds stress by introducing an eddy viscosity.

\section{Mesh independency of the local condition number function}
\label{sec:discretization}
We present the numerical discretization of the proposed local condition number on a CFD mesh and show that the discretized local condition number is mesh-independent. First, the function norms $\Cnorm{\cdot}$ are approximated  in vector norms $\Vnorm{\cdot}$ through numerical integration on a CFD mesh of $n$ cells. This is derived as follows:
\begin{linenomath}
\begin{align}
\Cnorm{G_{\bm{x}'}} & = \sqrt{\int_\Omega | G(\bm{x}'; \ms{\xi})|^2 \, 
d \ms{\xi} } \label{eq:fn-defition} \\
& \approx \sqrt{ \sum_{i=1}^n \left( [\mb{r}_{j,i}]^2 \, \Delta V_i \right) } \\
& = \sqrt{ \sum_{i=1}^n \left([\mb{r}_{j,i}] \, \sqrt{[\Delta V_i]}\right)^2} \\
& = \Vnorm{\mb{r}_j \sqrt{[\Delta \mb{V}]}}
=  \Vnorm{\tilde{\mb{r}}_j}
\end{align}
\end{linenomath}
with $\tilde{\mb{r}}_j = \mb{r}_j \sqrt{\Delta \mb{V}}$ and $\approx$ indicating numerical discretization of the integral involved in the function norm in Eq.~(\ref{eq:fn-defition}); $\Delta V_i$ is the volume for the $i$-th cell; $\Delta \mb{V}$ is the $n$-vector consisting of volumes of cells in the mesh; $\mb{r}_{j}$ is the $j$-th row of the matrix $\mathsf{A}^{-1}$, with $\mathsf{A}$ being the discretization of the operator $\mathcal{L}$ as seen in Eq.~(\ref{eq:L-definition}). The numbering implies that the location $\mb{x}'$ is the coordinate of the $j$-th cell. 

\begin{figure}
  \centering
      \includegraphics[width=0.5\textwidth]{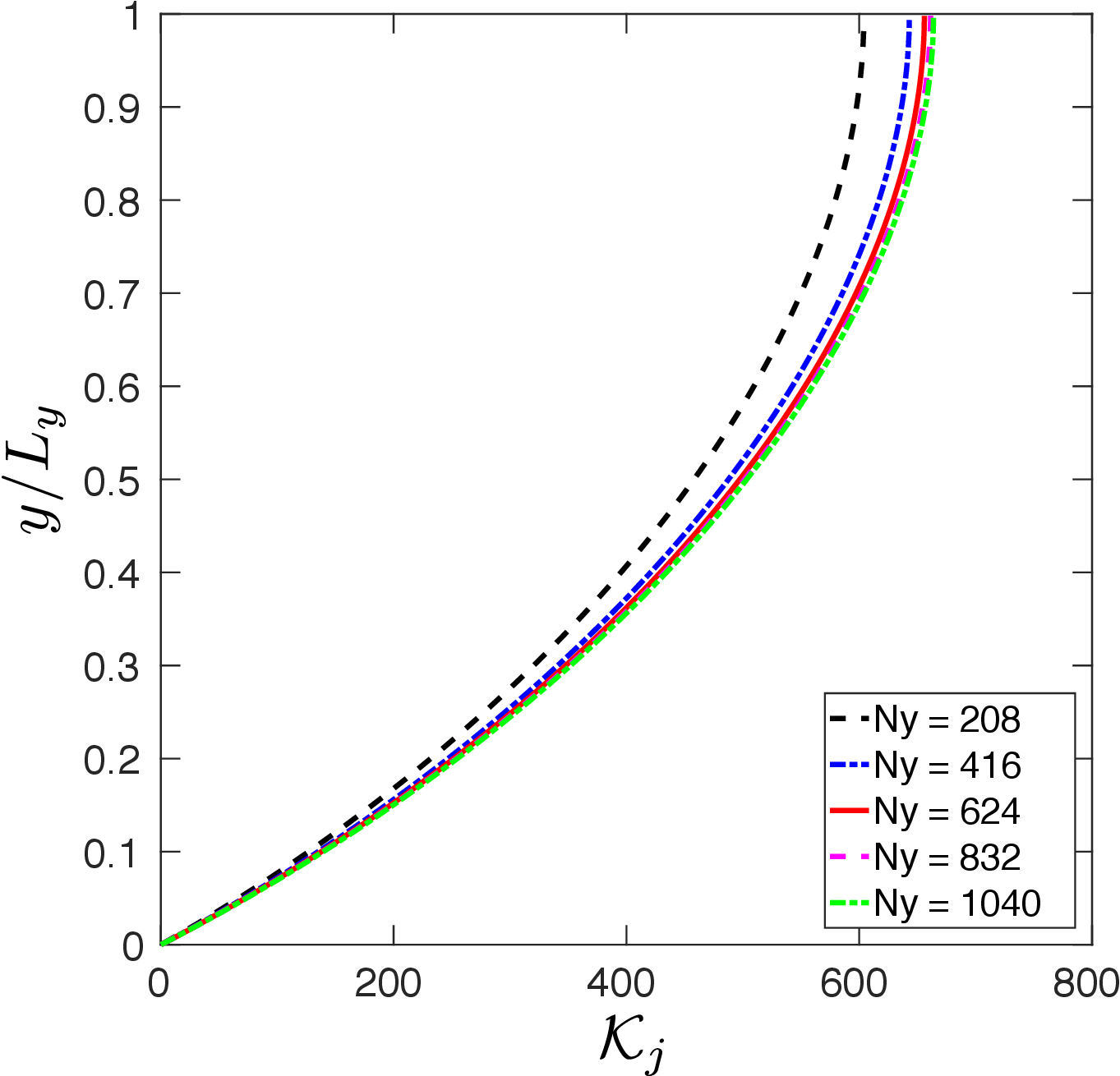}
    \caption{Mesh convergence study of local condition number $\mathcal{K}_j$ at $Re=5200$ by using explicit treatment with fixed Reynolds stress, i.e., the fixed DNS Reynolds stress is substituted into RANS equations.}
  \label{fig:mesh-converge}
\end{figure}

Similarly, the function norm of the Reynolds stress divergence is approximated on the CFD mesh as follows:
\begin{linenomath}
\begin{equation}
\Cnorm{\nabla \cdot \ms{\tau}} 
\approx  \Vnorm{[\nabla \cdot \ms{\tau}] \sqrt{[\Delta \mb{V}]}} 
\end{equation}
\end{linenomath}
\noindent
It is clear that the function norm $\Cnorm{G_{\mb{x}'}}$ is mesh-independent as its definition does not involve any discretization mesh (quadrature), so its numerical approximation 
$\Vnorm{\mb{r}_j \sqrt{[\Delta \mb{V}]}}$ should also be mesh independent on a sufficiently fine mesh.
In the same way, both the function norm $\Cnorm{\nabla \cdot \ms{\tau}}$ and its numerical approximation $\Vnorm{[\nabla \cdot \ms{\tau}] \sqrt{[\Delta \mb{V}]}}$ are mesh independent. The mesh independency can be further verified in the special case, where the divergence field $\nabla \cdot \ms{\tau}$ is a nonzero constant $\beta$ and the mesh consists of $n$ uniformly sized cells. In this case we have $\Delta V = \frac{|\Omega|}{n}$, where $|\Omega|$ is the total volume of the computational domain $\Omega$ (independent of the discretization mesh).  Therefore, the vector norm, which is a numerical approximation of the function form, is as follows:
\begin{linenomath}
\begin{align}
\Vnorm{[\nabla \cdot \ms{\tau}] \sqrt{[\Delta \mb{V}]}}  = &  
\sqrt{\sum_{i=1}^n  \left( \beta \sqrt{\frac{|\Omega|}{n}} \right)^2} \\
&   
= \sqrt{n \; \beta^2 \frac{|\Omega|}{n}} \\
&  
= \beta \sqrt{|\Omega|},
\end{align}
\end{linenomath}
which is clearly independent of the number of cells in the mesh. In order to complement and validate the derivations above, a numerical study of mesh convergence is presented in Fig.~\ref{fig:mesh-converge}, where the local condition number $\mathcal{K}_j$ at $Re=5200$ is calculated by using different mesh resolutions. It can be seen that asymptotical convergence of local condition number $\mathcal{K}_j$ can be achieved by gradually refining the mesh resolution.  Moreover, even on a coarser mesh the overall magnitude and spatial distribution of the condition number $\mathcal{K}_j$ are correctly captured.

\section{Explicit treatment of Reynolds stress with dependence on strain rate}
\label{sec:app-explicit-dependence}
\begin{figure}
  \centering
  \subfloat[Relative error]{\includegraphics[height=0.35\textwidth]{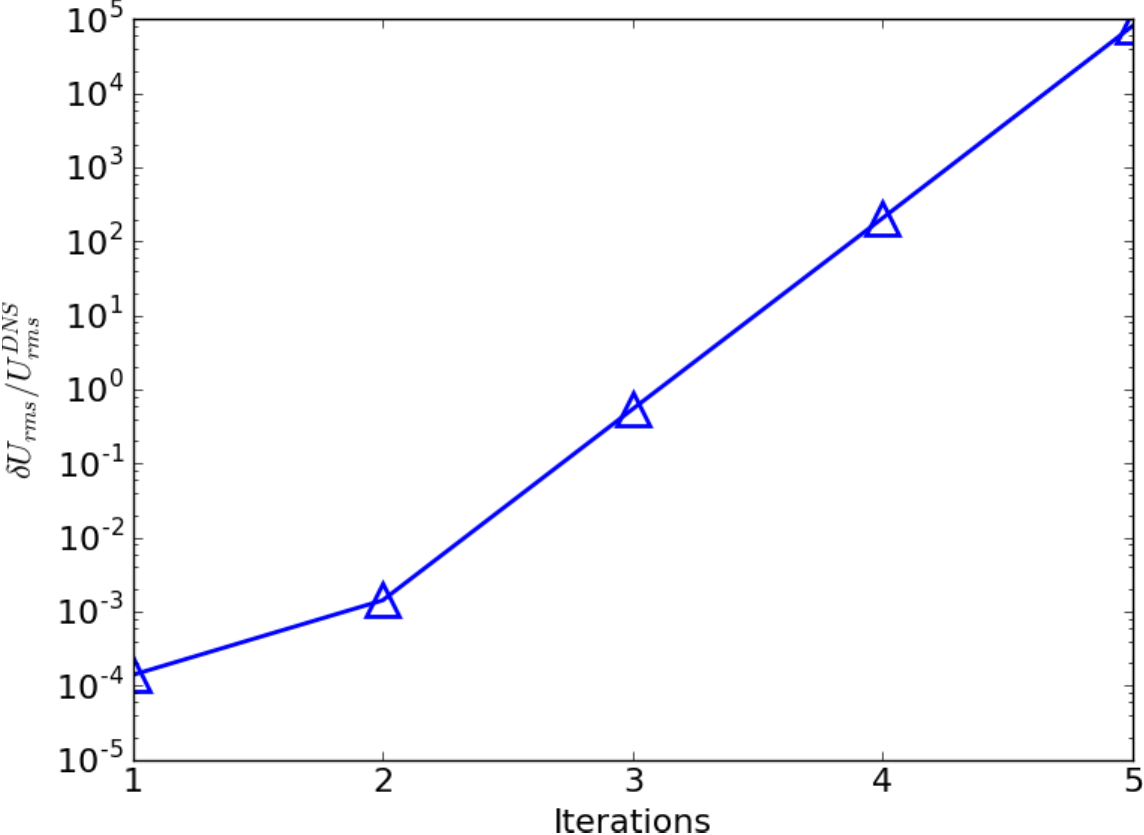}}\hspace{0.5em}
  \subfloat[Condition number]{\includegraphics[height=0.35\textwidth]{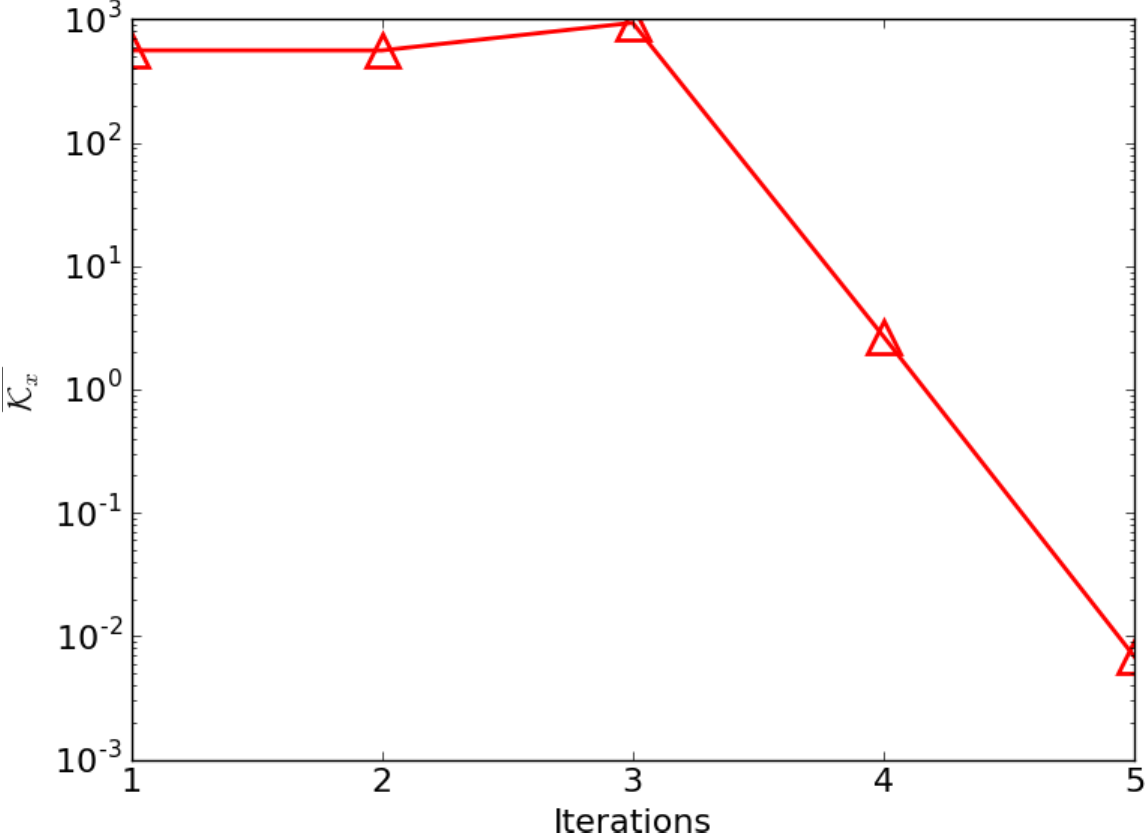}}
    \caption{The error propagation analysis of solving stream-wise velocity iteratively by using \textbf{explicit treatment} of Reynolds stress that depends on the strain rate, including (a) relative error of mean velocity and (b) volume-averaged local condition number.}
  \label{fig:coupled-Exp-Ux}
\end{figure}

The conditioning of RANS equations with explicit treatment of Reynolds stress has been discussed in Section~\ref{sec:explicit-treatment}, where the Reynolds stress is obtained from DNS database and is fixed when solving RANS equations. If the dependence of Reynolds stress on strain rate is considered, it can be seen in Fig.~\ref{fig:coupled-Exp-Ux} that the conditioning of RANS equations in the first iteration is the same as the conditioning with fixed Reynolds stress as shown in Fig.~\ref{fig:frozen-Ux}. It is because the detailed explicit treatment of Reynolds stress does not influence the model conditioning at a given state. Here we further show that the explicit coupling between Reynolds stress and mean velocity during the iterations can gradually amplify the errors and lead to divergence. Such explicit coupling is often used in the Reynolds stress transport models (RSTM). Specifically, the Reynolds stress is obtained by solving its transport equations with the mean velocity field at the previous iteration step. In the following, we use a simplified example with an iterative solver as shown in Algorithm~\ref{alg:exp-model} to illustrate the convergence issue of Reynolds stress transport models. In addition, we demonstrate that the proposed local condition number can be used to detect and explain the corresponding ill-conditioning issue during iterations.

The Reynolds stress at $i^{\textrm{th}}$ iteration step is explicitly treated by using DNS data according to Algorithm~\ref{alg:exp-model}. Unlike the data-driven Reynolds stress modeling as shown in Algorithm~\ref{alg:reynolds-stress-model}, this simplified explicit Reynolds stress treatment allows the update of Reynolds stress at each iteration based on the solved mean velocity field at the previous iteration step. Compared to the implicit treatment of Reynolds stress as shown in Algorithm~\ref{alg:eddy-viscosity-model}, the only difference of this explicit treatment is the computing of Reynolds stress with the mean velocity at the previous iteration step, which is indicated by the superscript $i-1$ at line 3 of Algorithm~\ref{alg:exp-model}.

\vspace{1em}
\begin{minipage}{0.95\textwidth}
\begin{algorithm}[H]
Compute optimal eddy viscosity $\nu_t^m$ from DNS Reynolds stresses based on Eq.~(\ref{eq:nut-definition}) \\
 \For{\textrm{each iteration step} $i=1,2,...,N$}{
  Compute the Reynolds stress: $\bm{\tau}^{(i)} = \nu_t^m \left(\nabla \overline{u}^{(i-1)}+(\nabla \overline{u}^{(i-1)})^T\right) + \bm{\tau}^{\perp, DNS}$\
  Solve the RANS equations: $\mathcal{N}\left(\overline{u}^{(i)}\right) = \nabla \cdot \bm{\tau}^{(i)}-\nabla p$ to obtain $\overline{u}^{(i)}$
 }
  \caption{\emph{Explicit} treatment of Reynolds stress that depends on strain rate among iterations}
  \label{alg:exp-model}
\end{algorithm}
\end{minipage}
\vspace{1em}

The errors of solved mean velocity field by using explicit treatment of Reynolds stress is presented in Fig.~\ref{fig:coupled-Exp-Ux}a. The DNS mean velocity is used as the initial field in RANS simulations, and thus the initial value of $\delta U_{rms}/U^\text{DNS}_{rms}$ is small. However, the value of $\delta U_{rms}/U^\text{DNS}_{rms}$ increases rapidly during the first several iteration steps. It demonstrates that the conditioning issue within each iteration can lead to error amplification, i.e., a small error in the modeled Reynolds stress can lead to noticeable errors in the solved mean velocity field and thus influence the modeled Reynolds stress in the next iteration step. Due to such coupling of error amplification, even a small error of modeled Reynolds stress can lead to divergence of simulation eventually. It can be seen in Fig.~\ref{fig:coupled-Exp-Ux}b that the volume-averaged local condition number is at $O(10^2)$ within the first three iteration steps, explaining the rapid growth of error in the solved mean velocity. The error of the solved mean velocity grows rapidly and eventually leads to divergence of the simulation. Therefore, the solved mean velocity is not presented in this work since a converged solution was not achieved. Therefore, the proposed local condition number is still of importance in traditional RANS modeling since it provides a quantitative assessment of model conditioning at every iteration step. The divergence of the mean velocity is because the Reynolds stress is updated according to the mean velocity at each iteration step. With the rapid increased error in the mean velocity at the first several iteration steps, the error in the Reynolds stress also increases accordingly. The decrease of the condition number as shown in Fig.~\ref{fig:coupled-Exp-Ux}b does not guarantee the decrease of the error in the mean velocity in such a scenario, considering that the error of mean velocity is also influenced by the error in the modeled Reynolds stress. Therefore, the small condition number needs to be interpreted with cautious when the source term in RANS equations changes during the simulation. With large condition number in some previous iteration steps, it is possible that the error of the modeled Reynolds stress becomes large and the mean velocity error keeps increasing even though the condition number decreases.

\section{Discrepancies in velocities obtained with different treatments}
\label{sec:app-discuss-paradox}
It was shown in Fig.~\ref{fig:U-comp-nut} that the solved mean velocity can be significantly different depending on whether the DNS Reynolds stress used to solve Eq.~(\ref{eq:ns-concise-N}) is treated explicitly or implicitly.
In other words, solving the following two equations
\begin{linenomath}
\begin{align}
\mathcal{L}(\bm{u}^\text{exp}) & = \nabla \cdot \bm{\tau}^\text{exp} -\nabla p \qquad
\text{and}  \label{eq:exp-u} \\
\mathcal{L}(\bm{u}^\text{imp}) & = \nabla \cdot \bm{\tau}^\text{imp}  -\nabla p
\label{eq:imp-u}
\end{align}
\end{linenomath}
yields drastically different velocities. The superscript $\text{imp}$ indicates the implicit treatment of Reynolds stress and $\text{exp}$ denotes the explicit treatment, i.e., $\bm{\tau}^\text{imp}=\nu_t\left(\nabla\bm{u}^\text{imp} +(\nabla \bm{u}^\text{imp})^T\right)+\bm{\tau}^{\perp,DNS}$ and $\bm{\tau}^\text{exp}=\bm{\tau}^\text{DNS}$. This finding apparently contradicts the common understanding in traditional CFD practice that the \emph{converged solution} of the mean velocity should be the same regardless of how the Reynolds stress is treated. Indeed, the Reynolds stresses used in the two formulations in Eqs.~(\ref{eq:exp-u}) and~(\ref{eq:imp-u}) are approximately equal, since $\nu_t\left(\nabla\bm{u}^\text{imp} +(\nabla \bm{u}^\text{imp})^T\right)+\bm{\tau}^{\perp,DNS} \approx \bm{\tau}^\text{DNS}$. More precisely, the difference between $\bm{u}^\text{imp}$ and $\bm{u}^\text{DNS}$ is about $0.1\%$, and the difference between $\bm{\tau}^\text{imp}$ and $\bm{\tau}^\text{exp}$ should be at the similar level. However, the condition number with regard to the nonlinear differential operator $\mathcal{L}$ is large for the flows at high Reynolds numbers, and thus a small difference between $\bm{\tau}^\text{imp}$ and $\bm{\tau}^\text{exp}$ can lead to a large difference between the solved mean velocities $\bm{u}^\text{imp}$ and $\bm{u}^\text{exp}$.

In addition, the better solution of $\bm{u}^\text{imp}$ with implicit treatment of Reynolds stress can be explained by the improved model conditioning, i.e., the condition number is smaller with regard to the linear differential operator $\widetilde{\mathcal{L}}=\mathcal{L}-\nu_t^m \nabla^2$ for the implicit treatment of Reynolds stress. Specifically, we first define an optimal Reynolds stress $\bm{\tau}^{op}$ that can lead to $\bm{u}^\text{DNS}$ by solving RANS equations:
\begin{linenomath}
\begin{equation}
    \mathcal{L}(\bm{u}^\text{DNS})=\nabla \cdot \bm{\tau}^{op} - \nabla p
\end{equation}
\end{linenomath}
where $\bm{\tau}^{op}$ denotes the true Reynolds stress that provides $\bm{u}^\text{DNS}$ by solving RANS equations. The errors $\|\bm{\tau}^\text{DNS}-\bm{\tau}^{op}\|$ and $\|\bm{\tau}^{\perp,DNS}-\bm{\tau}^{\perp,op}\|$ are of the same order of magnitude. Therefore, $\|\bm{u}^\text{imp}-\bm{u}^\text{DNS}\|$ is smaller than $\|\bm{u}^\text{exp}-\bm{u}^\text{DNS}\|$ due to the smaller sensitivity of solving mean velocity by using the implicit treatment of Reynolds stress.

\section*{Notation}
We use $\disc{\phi}$ to indicate the $n$-vector obtained by discretizing the field $\phi$ on the mesh, where
$n$ is number of cells/grid points. $\|\disc{\phi}\|$ denotes the norm of vector $\disc{\phi}$
resulted from the discretization. Since the norm is always taken on discretized $n$-vector, we
abbreviated $\|\disc{\phi}\|$ as $\|\phi\|$ without ambiguity. 
When mentioning function norm and $n$-vector norm simultaneously, we used $\Cnorm{\cdot}$ and $\Vnorm{\cdot}$ to distinguish them, with $\Omega$ denoting the domain on which the norm is defined.

\begin{tabbing}
  XXXX \= \kill
  $\bm{u}$ \> mean velocity field \\
  $\mathbf{U}$ \> discretized mean velocity ($n$-vector) \\
  $\bm{\tau}$ \>  Reynolds stress tensor \\
  $\mathbf{S}$ \> rate-of-strain tensor \\
  $\bm{b}$ \> imbalance vector between Reynolds stress divergence and pressure gradient \\
  $\mathsf{A}$ \> $n \times n$ coefficients matrix of discretized RANS equations \\
  $\mathcal{N}$ \> non-linear differential operator \\
  $\mathcal{L}$ \> linear differential operator \\
  $ G $ \>  Green's function corresponding to $\mathcal{L}$ \\
  $\mathcal{K}_\mathsf{A}$ \> condition number of matrix $\mathsf{A}$ \\
  $\overline{\alpha}$ \> ratio between Reynolds stress divergence and the total source term \\
  $\mathcal{K}_\tau$ \> matrix-norm condition number associated with Reynolds stress perturbation \\
  $ \mathcal{K}_j $ \> local condition number vector approximated on a CFD mesh ($n$-vector) \\
  $ \overline{\mathcal{K}}_x $ \> volume-averaged local condition number (scalar) \\
  $ \delta U_{rms} $ \> volume-averaged root-mean-squared error of solved mean velocity \\
    $ U_{rms}^\text{DNS} $ \> volume-averaged root-mean-squared DNS mean velocity \\
  $ \perp $ \> superscript indicating the non-linear part of Reynolds stress \\
   \end{tabbing}

\bibliographystyle{jfm}

\end{document}